\documentclass[structabstract]{aa}

\usepackage{epsfig}
\usepackage{amssymb,amsmath}
\usepackage{txfonts}

\usepackage{natbib}
\bibpunct{(}{)}{;}{a}{}{,} 

\begin{document}

\title{Comparison of the scintillation noise above different observatories measured with MASS instruments }
\author{V.~Kornilov\inst{1} \and M.~Sarazin\inst{2} \and A.Tokovinin\inst{3} \and T.~Travouillon\inst{4} \and O.~Voziakova\inst{1} }

\institute{Lomonosov Moscow State University, Sternberg Astronomical Institute, Universitetsky pr-t 13, Moscow, Russia
    \and European Southern Observatory, Karl-Schwarzschild-Strasse 2, D-85748, Garching, Germany
    \and Cerro Tololo Inter-American Observatory, Casilla 603, La Serena, Chile
    \and Thirty Meter Telescope Observatory Corporation, 1111 South Arroyo Parkway, Suite 200, Pasadena, CA 91105, USA}

\date{Accepted: August 13, 2012}

\abstract{}
{Scintillation noise is a major limitation of ground base photometric precision.}
{An extensive dataset of stellar scintillation collected at 11 astronomical sites world-wide with MASS instruments was used to  estimate the scintillation noise of large telescopes in the case of fast photometry and traditional long-exposure regime.}
{Statistical distributions of the corresponding parameters are given. The scintillation noise is mostly determined by turbulence and wind in the upper atmosphere and comparable at all sites, with slightly smaller values at Mauna Kea and largest noise at Tolonchar in Chile. We show that the classical Young's formula under-estimates the scintillation noise.The temporal variations of the scintillation noise are also similar at all sites, showing short-term variability at time scales of 1 -- 2 hours and slower variations, including marked seasonal trends (stronger scintillation and less clear sky during local winter). Some correlation was found between nearby observatories.}

\keywords{ Techniques: photometric -- Site testing -- Atmospheric effects }

\authorrunning{V.\,Kornilov et al}
\titlerunning{Comparison of the scintillation noise above different observatories}

\maketitle

\section{Introduction}

One of the main characteristics of astronomical objects is their brightness in different spectral bands. The standard precision of ground-based photometry is adequate in most cases, but a number of astronomical problems require an even greater precision \citep{Heasley1996,Everett2001}. One of the fundamental factors limiting the precision of ground-based photometry is the stellar scintillation occurring in the atmosphere as a result of its turbulent nature.

The fluctuations of the refractive index cause phase distortion in a plane light wave passing through the atmosphere to an entrance aperture of the telescope. As the wave propagates, the phase distortions lead to a redistribution of the amplitudes between different parts of the wavefront. Averaging within the aperture reduces the fluctuations caused by this mechanism, but does not eliminate them completely. In photometric practice, this effect is considered as an additional source of error. The scintillation noise expressed in stellar magnitudes does not depend on the object's brightness, therefore it cannot be reduced by observing brighter stars.

Scintillation noise in high-precision and fast photometry has been studied for quite a long time \citep{Young1967,Young1969,Dravins1997a}. The large number of methods proposed for reducing scintillation noise \citep{Heasley1996,Dravins1998,Gilliland1993,Osborn2010} shows that there is no perfect solution and that further work is needed. Moreover, this same problem is often presented as an argument to do precise photometry from space.

Stellar scintillation is of interest not only to photometry. This phenomenon is a powerful tool for remote sensing of optical turbulence (OT) in the atmosphere. The mechanisms of scintillation generation and its characteristics are well studied theoretically and experimentally because they are closely related to the most important characteristic of the OT above astronomical sites, the seeing.

Unlike the seeing, scintillation received relatively little attention in the astro-climatic work; the error budget of high-precision photometry is still evaluated using the data from \citep{Young1967} or other disparate estimates. Only recently such studies have been conducted in the general context of the characterization of optical turbulence above different astronomical observatories and prospective sites \citep{Kenyon2006,2011AstL}. These data are useful for comparing sites in the context of photometry. Moreover, the scintillation noise of long-exposure photometric measurements depends not only on the intensity of high-altitude OT, but on the wind speed at altitudes above the tropopause \citep{2011AstL}, which is important for the global dynamics of the atmosphere.

This paper presents scintillation noise measurements with the MASS instrument at observatories situated in different geographical areas. Sect.~\ref{sec:theory} recalls the theoretical description of the scintillation in three basic measurement regimes. In Sect.~\ref{sec:decomposition} the method used to estimate relevant parameters from the raw data is described. The next Section describes the sites studied here, the original data, and the procedure for calculating the scintillation noise. Results and comparative analysis are presented in Sect.~\ref{sec:results_s3}, Sect.~\ref{sec:temporal} describes the temporal variability of scintillation noise, and the final section is a discussion and comparison with other available data.

\section{Basic relations}

\subsection{Theoretical background}
\label{sec:theory}

Scintillation is characterized by the variance of relative fluctuations of the flux $I$ passing through a receiving aperture, the so-called scintillation index $s^2$:
\begin{equation}
s^2 = \langle (I-\langle I \rangle)^2 \rangle/\langle I \rangle^2.
\end{equation}

In the approximation of weak perturbations, the total scintillation index is the sum of scintillation produced by independent turbulent layers:
\begin{equation}
s^2 = \int_0^\infty C_n^2(z) W(z)\,{\rm d}z,
\label{eq:is}
\end{equation}
where $W(z)$ is a weighting function which depends on the size and shape of the receiving aperture and does not depend on the distribution of the structural refractive-index coefficient $C_n^2(z)$. The weighting function has a simple physical meaning, it equals the scintillation index generated by a layer of unit intensity $C_n^2 {\rm d}z$ located at a distance $z$.

In the general case of non-zero exposure, the weighting function also depends on the exposure time and on the wind speed at altitude $h$. The altitude and distance to the layer are trivially related as $h = z\,\cos \gamma$, where $\gamma$ is the zenith angle of the observed object. Theoretical descriptions of the scintillation can be found, for example, in \citet{Young1967,Roddier1981,Dravins1997a}.

Using the Taylor's frozen-flow hypothesis \citep{Taylor1938}, the wavefront evolution during exposure time $\tau$ for wind speed $w(z)$ is reduced to a simple translation by $w(z) \tau$. Invoking the Kolmogorov spectrum of refractive index perturbations, the final expression for the weighting function is \citep{Tokovinin2002b,2011AstL}:
\begin{equation}
W(z,w,\tau) = 9.62\int_0^\infty {\rm d}f \, f^{-8/3} S(z,f) A(f) A_s(w\tau,f).
\label{eq:wdef}
\end{equation}

Here, the integration is performed over the modulus of the spatial frequency $f$, assuming that the functions in the integrand are either axisymmetric or are already averaged over the polar angle. The function $A(f)$ is the aperture filter, which takes into account spatial averaging by the aperture. $A_s(w, \tau, f)$ is the spectral filter of the wind translation which describes the temporal averaging, and $S(z,f)$ is the Fresnel spectral filter which describes the generation of amplitude distortions in the propagation of the wavefront.

For a circular aperture, the aperture filter is $A(f) = [2\,J_1(\pi Df)/\pi Df]^2$. The Fresnel filter $S(z, f)$ depends on the wavelength $\lambda$ of the detected radiation and can be calculated for any spectral energy distribution \citep{Tokovinin2003}. In the case of monochromatic radiation $S(z, f) = [\sin (\pi \lambda z f^2)/\lambda]^2$. The characteristic spatial scale of the filter is the Fresnel radius $r_\mathrm{F} = (\lambda z)^{1/2}$.

The wind-translation filter is not axisymmetric, but after averaging over the polar angle it is represented as $A_s(w, \tau, f) = \mathcal T_1(f\tau w)$, where the function $\mathcal T_1(\xi)$ can be expressed in terms of special functions \citep{Tokovinin2002b,2011AstL}. The important features of this function are its asymptotes for small $\xi$: $\mathcal T_1(\xi) \approx 1 - \pi^2\xi^2/6$ and for large $\xi$: $\mathcal T_1(\xi) \approx 1/\pi \xi$.

It is difficult to analyse the expression (\ref{eq:wdef}) in general form because it depends on many parameters. However, for certain relationships between the parameters the expression is considerably simplified. First of all, it is the definition of the small $D \ll r_\mathrm{F}$ and large $D \gg r_\mathrm{F}$ aperture regimes. In the first case we can set $A(f) \equiv 1$, in the second case we simplify the Fresnel filter by replacing the sine with its argument \citep{Roddier1981}.

We can also identify two limiting cases by the value of the wind translation. At very short (zero) exposures $w \tau \ll \min(D,r_\mathrm{F})$, the filter $A_s(w \tau, f) \equiv 1$. In the opposite case of long exposure $w \tau \gg \max (D, r_\mathrm{F})$, the $ A_s(w \tau, f) = 1/\pi w \tau f$.

These limiting situations are generally accepted, although in photometric practice, the small aperture approximation is not as interesting. In this paper, we will consider the evaluation of scintillation noise on a large telescope in the regimes of short (SE) and long (LE) exposures.

The scintillation index in a large telescope for short exposures is expressed by the well-known formula \citep[see, e.g.,][]{Roddier1981}:
\begin{equation}
s^2_\mathrm{S} = 17.34\,D^{-7/3}\int_A C_n^2(z)\,z^2 {\rm d}z,
\label{eq:short_s2}
\end{equation}
since in this case the weighting function is $W(z) = 17.34\,D^{-7/3}\,z^2$. The weighting function does not depend on the wavelength owing to the achromatism of the scintillation in a large telescope. This case is typical for fast photometry. The dependence on the zenith distance $\gamma$ can be obtained from trivial geometric considerations: $s^2_\mathrm{S}\propto M_\mathrm{z}^3$, where $M_z = \sec \gamma$. This formula ignores the central obscuration of astronomical telescopes, important in the short-exposure regime \citep{Young1967,Dravins1998}.

Nevertheless, in conventional astronomical photometry typical exposures of tens of seconds and longer are used. In this case, the condition $w\tau\gg D$ for long exposure is satisfied for typical wind speeds in the upper atmosphere, and \citep{2011AstL}
\begin{equation}
s^2_\mathrm{L} = 10.66\,D^{-4/3}\tau^{-1} \int_A \frac{C_n^2(z)\,z^2}{w(z)}{\rm d}z.
\label{eq:long_s3}
\end{equation}

Naturally, the wavefront is translated by the wind's component perpendicular to the line of sight. Given that, on average, the wind vector is directed horizontally, this effect also depends on the air mass $M_\mathrm{z}$ and on the azimuth of the wind with respect to the observed star \citep{Young1969}. The dependence of the scintillation on the air mass therefore varies from $s^2_\mathrm{L}\propto M_\mathrm{z}^3$ for the case of transverse wind up to $s^2_\mathrm{L}\propto M_\mathrm{z}^4$ in the case of longitudinal wind directed along the object azimuth. In the long-exposure regime the scintillation is little affected by the central obscuration \citep{Young1967}, which has to be taken into account only in the extreme case of a ring-like aperture \citep{2012bMNRAS}.

Formulae (\ref{eq:short_s2}) and (\ref{eq:long_s3}) show that for a particular telescope and exposure time the scintillation index can be easily calculated if the integrals in these formulae are known. Using the notation introduced in \citet{Kenyon2006}:
\begin{equation}
S^2_2 = 17.34\int_A C_n^2(z)\,z^2 {\rm d}z = 17.34\,\mathcal M_2,
\label{eq:s2m2}
\end{equation}
for the SE regime, and
\begin{equation}
S^2_3 = 10.66\int_A \frac{C_n^2(z)\,z^2}{w(z)}{\rm d}z = 10.66\,\mathcal Y_2,
\label{eq:s3y2}
\end{equation}
for LE regime.

These {\em scintillation noise parameters} $S^2_2$ and $S^2_3$ characterize the power of scintillation noise under given atmospheric conditions and represent the scintillation index in a 1~m telescope (with 1~s exposure for the LE regime). The quantities $\mathcal M_2$ and $\mathcal Y_2$ are known as atmospheric moments of second degree \citep{2003MNRAS}.

\subsection{Measurement of scintillation noise parameters with MASS}
\label{sec:decomposition}

Multi-Aperture Scintillation Sensor (MASS) is an instrument for measuring OT profile from scintillation \citep{2003SPIE,2007bMNRAS}. Photons from a single bright star collected by a small telescope are detected with four photo-multipliers which sample intensity in the pupil with four concentric annular apertures A, B, C, and D. The smallest circular aperture A at the centre has a typical diameter of 2~cm, the largest annular aperture D has an outer diameter of $\sim 10$~cm. These diameters are determined by the optical magnification factor $k$, adjusted and measured for each instrument individually. Photon counts with a 1-ms exposure are processed statistically to derive 4 normal and 6 differential scintillation indices which are related to the OT through weighting functions (Eq.~\ref{eq:is}). Apart from the aperture geometry, these functions depend on the spectral energy distribution, i.e. on the spectral type of the star and on the instrument response. The functions $W(h)$ are computed by numerical integration of the basic expression (\ref{eq:wdef}). In the calculations, it is assumed that the measurement occurs with ``zero'' exposure, neglecting the wind translation during 1\,ms \citep{2003MNRAS,2011ExA}.

The measured scintillation indices are used not only for the restoration of the vertical profile of OT, but also for the evaluation of several atmospheric moments required for computing such integral characteristics of the atmosphere as the free atmosphere seeing, isoplanatic angle, effective altitude of the turbulence, etc. \citep{2003MNRAS,2007bMNRAS}. Among these moments, the second moment $\mathcal M_2$ is computed, so the evaluation of the parameter $S^2_2$ is a trivial task.

The method of finding the parameter $S^2_3$ has been proposed in \citet{2011AstL}. It relies on the fact that the scintillation index $s^2_\mathrm{L}$ for long exposure can be calculated from the mean fluxes measured with 1~s exposure which are stored in the MASS output files {\tt *.stm}. It should be mentioned that for the small MASS apertures the LE regime begins at $\tau \ll 1$\,s.

In this LE regime, the scintillation index (we call it LE-index to distinguish from the "fast" indices with 1\,ms exposure) can be expressed as
\begin{equation}
s^2_\mathrm{L} = \int_A \frac{C_n^2(z)}{w(z)\tau}\,U'(z) {\rm d}z,
\label{eq:longMASS}
\end{equation}
where $U'(z)$ are the MASS weighting functions for the LE regime. They are distinguished from the conventional weighting functions by the additional factor $(\pi f)^{-1}$ under the integral over the modulus of the spatial frequency. Typical behaviour of $U'(z)$ for the whole set of MASS indices is presented in \citet{2011AstL}, so we only note that the asymptotic dependencies of these functions are $U'(z) \propto z^{4/3}$ for infinitely small apertures, and $U'(z) \propto z^{2}$ for large apertures.

We estimate $S^2_3$ from the weighted sum (linear combination) of LE scintillation indices measured with MASS. The coefficients of this sum are adjusted to make the weighted sum of $U'(z)$ as close as possible to $10.66\,z^2$. Then, as can be seen by comparing Eq.~\ref{eq:s3y2} and Eq.~\ref{eq:longMASS}, the weighted sum will approximate $S^2_3$. Details of the procedure are given in appendix~\ref{sec:approxim}

\begin{figure}
\centering
\psfig{figure=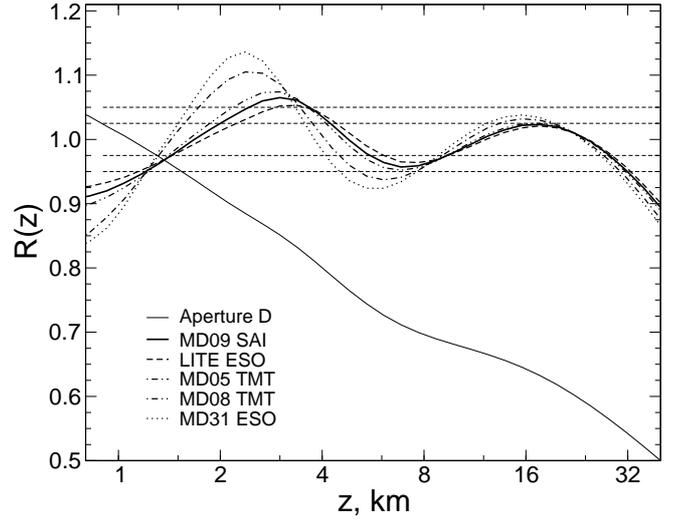,height=9cm,angle=-90}
\caption{The ratio $R(z)$ of the combined LE weighting functions to $10.66\,z^2$. Horizontal lines mark the $\pm 5$\% and $\pm 2.5$\% corridors. The thin solid line depicts this ratio for the D aperture.
\label{fig:appr}}
\end{figure}

Fig.~\ref{fig:appr} plots the relative error of such approximations for five typical MASS devices. It shows that for propagation distances between 4 and 32~km, the errors do not exceed $\pm 5\%$. The contribution of closest layers (below 1~km) is underestimated, but it is negligible due to the factor $z^2$. The impact of the distant (beyond 32~km) OT is underestimated by $\approx 10\%$ as well. Here we exclude the measurements made far from the zenith ($M_\mathrm{z} > 1.3$), so this under-estimation is relevant for altitudes above $25$~km where the OT has already low intensity.

This figure also shows the curve for the aperture D of the device with $k = 16.3$. It is evident that if we only use this aperture, the result is greatly underestimated. The approximation is therefore better for devices with larger apertures (larger magnification coefficient) and a higher spectral response in the blue.

The influence of the magnification coefficient on the results was verified a posteriori for the real data obtained with the device MD05 TMT (Armazones). When changing $k$ from 15.9 to 15.4 (by $\approx 3\%$), the median value of the parameter $S^2_3$ decreased by $4\%$, so that the requirement for the accuracy of the $k$ is not stronger than in the typical processing of MASS data \citep{2007bMNRAS}. The substitution of another spectral reaction curve with an effective wavelength smaller by $\approx 14$~nm changed the median by less than $0.2\%$. The dependence of the approximation on the star colour is negligible.

\section{Data processing}
\label{sec:dataproc}

\subsection{Data sets and site testing campaigns}

In the past decade, a large amount of scintillation data from MASS instruments has been accumulated. The measurements were performed by various projects in different geographical locations. Some of these campaigns focused on the site selection for future large telescopes. Other studies were methodological in nature, or supported already operating telescopes. The list of astronomical sites studied in this paper is listed in the Table~\ref{tab:sites}.

\begin{table*}
\caption{Astronomical sites and main characteristics of the site testing campaigns. Observatory coordinates, altitude above sea level $H$ and total measurement time $T$ are listed.\label{tab:sites}}
\centering
\begin{tabular}{lllllcr}
\hline\hline
Site & Project & Longitude & Latitude & $H$, m & Period & $T$, h \\
\hline
Armazones & TMT & $-04^\mathrm{h}40^\mathrm{m}44^\mathrm{s}$ & $-24\degr 34\arcmin 48\arcsec$ & 3\,064 & 11/2004 -- 05/2009 & 4\,406 \\ 
 & ESO & & & & 01/2010 -- 12/2011 & 2\,703 \\ 
La Chira & ESO & $-04^\mathrm{h}41^\mathrm{m}23^\mathrm{s}$ & $-24\degr 30\arcmin 20\arcsec$ & 2\,559 & 11/2006 -- 11/2007 & 888 \\ 
Mauna Kea & TMT & $-10^\mathrm{h}21^\mathrm{m}55^\mathrm{s}$ & $+19\degr 49\arcmin 31\arcsec$ & 4\,204 & 07/2005 -- 05/2008 & 2\,478 \\ 
Pach\'on & CTIO & $-04^\mathrm{h}42^\mathrm{m}56^\mathrm{s}$ & $-30\degr 14\arcmin 24\arcsec$ & 2\,738 & 11/2004 -- 02/2012 & 8\,642 \\ 
Paranal & ESO & $-04^\mathrm{h}41^\mathrm{m}36^\mathrm{s}$ & $-24\degr 37\arcmin 31\arcsec$ & 2\,635 & 09/2004 -- 12/2011 & 14\,122 \\ 
San Pedro M\'artir &TMT&$-07^\mathrm{h}41^\mathrm{m}51^\mathrm{s}$ & $+31\degr 02\arcmin 38\arcsec$ & 2\,800 & 10/2004 -- 08/2008 & 1\,642\\ 
Shatdzhatmaz & SAI & $+02^\mathrm{h}50^\mathrm{m}40^\mathrm{s}$ & $+43\degr 44\arcmin 12\arcsec$ & 2\,110 & 11/2007 -- 11/2011 & 3\,003 \\ 
Tolar & TMT & $-04^\mathrm{h}40^\mathrm{m}24^\mathrm{s}$ & $-21\degr 57\arcmin 50\arcsec$ & 2\,290 & 10/2004 -- 03/2006 & 1\,344 \\ 
Tololo & CTIO & $-04^\mathrm{h}43^\mathrm{m}15^\mathrm{s}$ & $-30\degr 09\arcmin 55\arcsec$ & 2\,215 & 04/2009 -- 04/2012 & 2\,474 \\ 
Tolonchar & TMT & $-04^\mathrm{h}31^\mathrm{m}54^\mathrm{s}$ & $-23\degr 56\arcmin 10\arcsec$ & 4\,480 & 01/2006 -- 07/2008 & 1\,378 \\ 
Ventarrones & ESO & $-04^\mathrm{h}40^\mathrm{m}50^\mathrm{s}$ & $-24\degr 23\arcmin 57\arcsec$ & 2\,837 & 01/2008 -- 02/2010 & 2\,180 \\ 
\hline\hline
\end{tabular}
\end{table*}

They include all sites studied by the TMT site testing program \citep{Schoeck2009}, representing two sites in the northern hemisphere: Mauna Kea (Hawaii) and San Pedro Martir (Mexico) and three mountains in Northern Chile: Cerro Tolar, Armazones and Tolonchar. The data from this campaign uses a similar setup on all sites using MASS/DIMM units attached to custom made telescopes capable of operating robotically within a large range of wind conditions.

Measurements on Cerro Armazones in Chile were first carried out by TMT. After this site was chosen for the E-ELT, monitoring was taken over in 2010 by ESO. Shorter samples at nearby sites Ventarrones and La Chira, studied in the frame of the E-ELT site characterization campaign \citep{Vernin2011}, have been added for the purpose temporal correlation analysis.

The robotic MASS-DIMM site monitor at Cerro Pach\'on is a shared facility between Gemini-S and SOAR telescopes, operational since 2004. Early results of turbulence monitoring at Pach\'on were reported in \citet{Tokovinin2006MN}. Here we use the much larger data set accumulated during 8 years.

\begin{figure}
\centering \psfig{figure=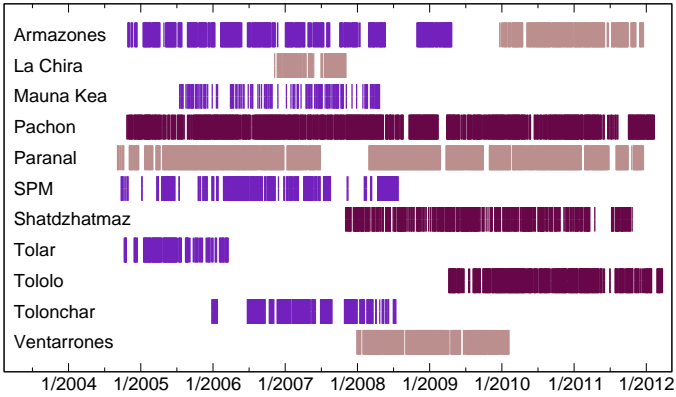,height=9.1cm,angle=-90}
\caption{Time distribution of the measurements. Indigo colored stripes are TMT campaigns, brown stripes are ESO campaigns, maroon stripes are CTIO and SAI campaigns.\label{fig:shart_time}}
\end{figure}

The longest series of observations were obtained at the Paranal and Pach\'on observatories. To control the stability of the results and for comparison with the two campaigns at Armazones, the Paranal data are also divided in two parts: 2004 -- 2009 (Paranal A) and 2010 -- 2011 (Paranal B). Since the measurements are carried out at Paranal in continuous mode, both samples are quite large. The time coverage of all campaigns is shown in Fig.~\ref{fig:shart_time}. The overall volume of data amounts to more than 40\,000 hours.

The only site in the eastern hemisphere is represented by data from the Mt.~Shatdzhatmaz in the Northern Caucasus \citep{2010MNRAS} where Sternberg Astronomical Institute (SAI) observatory is building its 2.5-m telescope. Note
that in operating observatories, the measurements with MASS device are continuing to provide operational information about OT in the atmosphere.

\subsection{Computation of LE scintillation indices}

For each of four MASS apertures, the LE scintillation index $s^2_\mathrm{L}$ is calculated as mean square of the difference between adjacent 1~s mean fluxes:
\begin{equation}
s^2_\mathrm{L} = \frac{1}{2(N-1)}\frac{1}{\bar F^2}\sum_{i=0}^{N-1}(F_i-F_{i+1})^2 - 0.001/\bar F,
\label{eq:sdif}
\end{equation}
where $\bar F$ is the average of the series of $N$ photon counts $\{F_i \}$ with 1\,s exposure. In the MASS output files, these counts are scaled to 1~ms exposure, which causes the factor of $10^{-3}$ in the last term describing the contribution of the photon noise. Use of the difference effectively suppresses the contribution of the flux variations on time intervals greater than 1~s, unrelated to the scintillation. It is easy to show that the scintillation is uncorrelated at long exposures; it follows from the relation $s^2(\tau) \propto 1/\tau$ in the LE regime.

Each estimate of the $s^2_\mathrm{L}$ uses all 1-s flux values during the MASS accumulation time. As a rule, this time is 1 minute, hence $N = 60$. For an unbiased estimate of the variance of the difference, a factor of $1/(N-1)$ is required; although the number of differences is also $N-1$, the mathematical expectation is known and equal to zero.

For fluxes registered with 1~s exposure, the photon counting statistics can be considered as exactly Poissonian. Any difference of the PMTs from an ideal detector does not produce errors exceeding 1\%, even in the case of extremely small indices. Non-linearity of the photon counts is corrected with a fixed dead time of 20\,ns, as typical for the MASS detectors.

In contrast to \citet{2011AstL}, the reduction to the zenith is performed after evaluation of the parameter $\tilde S^2_3$ from the LE indices measured at a certain air mass $M_\mathrm{z}$, by Eq.~(\ref{eq:s3sum}). Since we do not know the direction of high-altitude winds, two estimates of the scintillation parameter at zenith are calculated: the minimum $S_3^2 = \tilde S_3^2\,M_\mathrm{z}^{-4}$ for the longitudinal wind, and the maximum $S_3^2 = \tilde S_3^2\,M_\mathrm{z}^{-3}$ for the transverse wind (see Sect.~\ref{sec:theory}). Data obtained at air masses $M_\mathrm{z} > 1.3$ are not used to avoid the uncertainty in the reduction to the zenith caused by the unknown wind direction. In addition, as noted in Sect.~\ref{sec:decomposition}, with a large air mass our method underestimates the contribution of high-altitude turbulence to the parameters $S_3^2$ and $S_2^2$.

Filtering used to remove invalid data and the estimates of the resulting errors are considered in Appendix~A. The filtering procedure eliminates a small, though non-negligible, fraction of the data, on average about 10\%. In some campaigns, a significant part of measurements was carried out far enough from the zenith; in these cases, the proportion of the rejected measurements exceeds 20\%.

Although we took great care in filtering out the bad data, the final results remain almost the same when all the data are used. This is so because the data were collected by automatic monitors for a long time and therefore contain only a small number of faulty measurements.

\section{Results}
\label{sec:results_s3}

\subsection{The scintillation parameter $S_2^2$}

\begin{figure}
\centering \psfig{figure=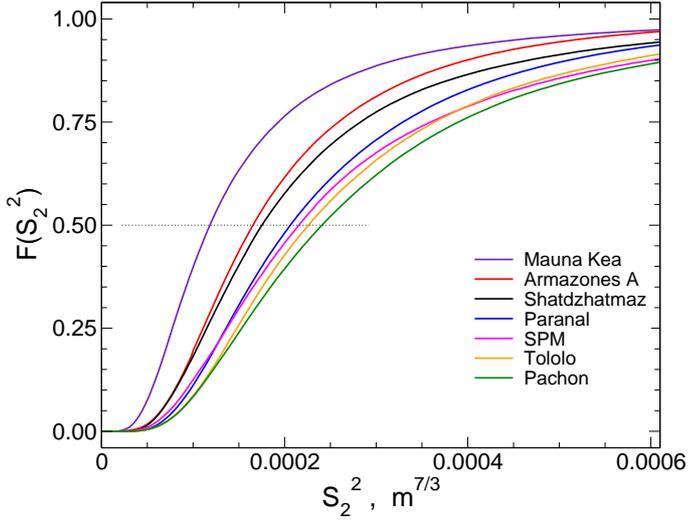,height=9cm,angle=-90}
\caption{Cumulative distributions of the $S_2^2$ parameter. The curves are labelled from left to right. Four sites are not plotted. \label{fig:s2-cumm}}
\end{figure}

As noted in Sect.~\ref{sec:decomposition}, the calculation of the parameter $S_2^2$ is made by rescaling the atmospheric moment $\mathcal M_2$ obtained in conventional MASS data processing and already reduced to the zenith (see Eq.~\ref{eq:s2m2}). Filtering  of the original data consists only in the choice of the points corresponding to the correct estimates of the parameter $S_3^2$. For some measurements, for various reasons, the $\mathcal M_2$ have not been obtained while the corresponding parameter $S_3^2$ has been successfully evaluated. These situations are reflected by the fact that in Table~\ref{tab:s2} the total time $T$ is somewhat smaller than in Table~\ref{tab:s3}.

Note that the algorithm of calculating $\mathcal M_2$ implemented in the early versions of the MASS software had some instability, which caused artifacts in the distribution and a systematic overestimation of the second moment. Therefore, in almost all cases we used the results of data reprocessing carried out later with the latest version of the software {\tt atmos-2.93} \citep{2011ExA}. For measurements on Tololo and Pach\'on we used the $\mathcal M_2$ calculated from the restored vertical profiles of the OT.

The cumulative distributions of the index $S_2^2$ (in $\mbox{m}^{7/3}$) are shown in Fig.~\ref{fig:s2-cumm}, and their characteristic points are given in Table~\ref{tab:s2}. Of all the distributions, the Mauna Kea site stands out, as expected in view of its high altitude and low turbulence in the upper atmosphere. The rest of the curves differ only in details, especially if we take into account the sampling effects for those campaigns with non-uniform seasonal coverage (we discuss this in the following Section) and some differences in the MASS devices. The fact that the Tololo and Pach\'on sites show a stronger scintillation noise is likely a result of evaluating $S_2^2$ at these sites from the OT profiles. Nevertheless, this systematic difference does not exceed 10\%.

The comparison of the Paranal~A and Paranal~B sub-samples shows that the period 2010 -- 2011 does not differ in scintillation power from the previous years. The same should be expected for the measurements at the Armazones summit due to its geographical proximity to the Paranal (30~km). However, the parameter $S_2^2$ for Armazones~B is higher by 15\% than for Armazones-A. Apparently, this discrepancy is caused by a significant difference in the geometry of the entrance apertures of the instruments used in both cases. A similar systematic difference is observed for the parameter of $S_3^2$, see Table~\ref{tab:s3}.

\begin{table}
\caption{Characteristic points of the distributions of the $S_2^2$ scintillation parameter. The units are $10^{-4}\mbox{ m}^{7/3}$. The estimates of relative error $E$ and the total  accumulated time $T$ are also listed. \label{tab:s2}}
\centering
\begin{tabular}{lrrrrrrrr}
\hline\hline
 & \multicolumn{3}{c}{$S_2^2$ quartiles } & $E$ & $T$, h\\
 & 25\% & 50\% & 75\% & & \\
\hline
Armazones A & 1.11 & 1.65 & 2.59 & 0.09 & 3810 \\
Armazones B & 1.26 & 1.94 & 3.13 & 0.08 & 1800 \\
La Chira & 1.60 & 2.42 & 3.93 & 0.08 & 520 \\
Mauna Kea & 0.77 & 1.18 & 1.93 & 0.09 & 2220 \\
Pach\'on & 1.53 & 2.41 & 3.90 & 0.09 & 4890 \\
Paranal A & 1.34 & 2.06 & 3.29 & 0.07 & 5630 \\
Paranal B & 1.38 & 2.08 & 3.30 & 0.07 & 3050 \\
S.~Pedro Martir & 1.38 & 2.18 & 3.68 & 0.09 & 1460 \\
Shatdzhatmaz & 1.15 & 1.75 & 2.83 & 0.06 & 2400 \\
Tolar & 1.29 & 1.86 & 2.86 & 0.09 & 1000 \\
Tololo & 1.48 & 2.26 & 3.64 & 0.10 & 1800 \\
Tolonchar & 1.34 & 2.14 & 3.64 & 0.09 & 940 \\
Ventarrones & 1.44 & 2.26 & 3.62 & 0.09 & 1430 \\
\hline\hline
\end{tabular}
\end{table}

The differential distributions are clearly asymmetric. For all the studied sites, the modes of the distributions coincide with the $23-26\%$ quantiles, so one can use the first quartile from Table~\ref{tab:s2} as the most probable value of $S_2^2$. The typical skewness $\gamma_1$ amounts to $2 - 3$ and the excess kurtosis is $\gamma_2 \sim 10$. Here we do not give more precise values, because adequate calculation of the higher moments of a statistical distribution requires careful removal of outliers and long-term trends.

The distributions of $\ln S_2^2$ are close to the normal distributions with slightly different means and widths. It is also noticeable that the asymmetry of the distributions are somewhat different for different observatories. Typical skewness and excess kurtosis of $\ln S_2^2$ are small, $\gamma_1 \sim 0.2 - 0.4$ and $\gamma_2 \sim 0 - 0.2$.

\subsection{The $S_3^2$ scintillation parameter}

The relative errors of $S_3^2$ estimates are quite large, $\sim 0.3 - 0.4$, widening somewhat the distributions. The significance of this effect was estimated by comparing the distributions of the initial results obtained with 60~s and with 240~s accumulation time (the latter are averages of four consecutive 60-s measures). As expected, the distribution of 240-s estimates is somewhat narrower than that of the 60-s estimates, but its median is larger by about 5\% because of the asymmetry.

\begin{figure}[h]
\centering \psfig{figure=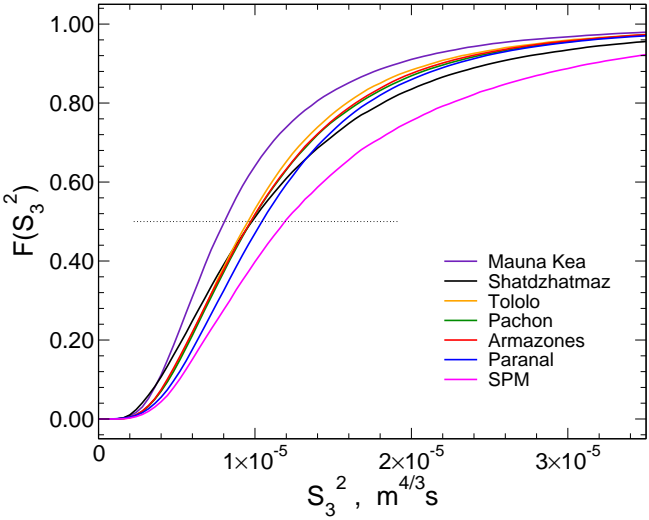,height=9cm,angle=-90}
\caption{Cumulative distributions of the $S_3^2$ parameter at seven sites. The curves for  different sites are labelled from left to right.\label{fig:s3distr}}
\end{figure}

In Fig.~\ref{fig:s3distr} the cumulative distributions of the 240~s estimates of $S^2_3$ are shown for all sites. For clarity, only the curves corresponding to the assumption of the longitudinal wind (i.e. the lower limit of $S_3^2$) are plotted. In Table~\ref{tab:s3} the characteristic points of the distributions are listed for both wind directions; they can be considered as lower and upper limits of the $S^2_3$ parameter.

\begin{table}
\caption{Characteristic points of the distributions of the $S_3^2$ parameter averaged over 4 minutes. The units are $10^{-5}\mbox{ m}^{4/3}\,\mbox{s}$. The estimates of relative error $E$ and the total accumulated time $T$ are also listed.\label{tab:s3}}
\centering
\tabcolsep=4pt
\begin{tabular}{@{}lrrrrrrrr@{}}
\hline\hline
& \multicolumn{3}{c}{Longitudinal wind} & \multicolumn{3}{c}{Transversal wind} & $E$ & $T$, h\\
$S_3^2$ quartiles & 25\% & 50\% & 75\% & 25\% & 50\% & 75\% & & \\
\hline
Armazones A & 0.64 & 0.97 & 1.48 & 0.71 & 1.09 & 1.66 & 0.17 & 3820 \\
Armazones B & 0.73 & 1.14 & 1.83 & 0.82 & 1.28 & 2.07 & 0.17 & 1880 \\
La Chira & 0.82 & 1.21 & 1.81 & 0.92 & 1.35 & 2.06 & 0.16 & 530 \\
Mauna Kea & 0.54 & 0.80 & 1.23 & 0.59 & 0.88 & 1.34 & 0.18 & 2270 \\
Pach\'on & 0.65 & 0.98 & 1.49 & 0.76 & 1.14 & 1.74 & 0.17 & 6800 \\
Paranal A & 0.70 & 1.04 & 1.55 & 0.80 & 1.20 & 1.78 & 0.16 & 5650 \\
Paranal B & 0.73 & 1.08 & 1.63 & 0.82 & 1.22 & 1.83 & 0.16 & 3050 \\
S.\,Pedro M\'artir & 0.76 & 1.19 & 1.97 & 0.82 & 1.30 & 2.15 & 0.16 & 1470 \\
Shatdzhatmaz & 0.60 & 0.98 & 1.60 & 0.66 & 1.07 & 1.76 & 0.16 & 2760 \\
Tolar & 0.78 & 1.12 & 1.59 & 0.89 & 1.28 & 1.85 & 0.16 & 1005 \\
Tololo & 0.65 & 0.97 & 1.45 & 0.76 & 1.13 & 1.70 & 0.17 & 1890 \\
Tolonchar & 0.89 & 1.40 & 2.25 & 1.00 & 1.61 & 2.59 & 0.17 & 1000 \\
Ventarrones & 0.70 & 1.07 & 1.62 & 0.78 & 1.20 & 1.81 & 0.17 & 1440 \\
\hline\hline
\end{tabular}
\end{table}

Both the curves in Fig.~\ref{fig:s3distr} and the data in Table~\ref{tab:s3} show that the distributions of the $S_3^2$ at various observatories differ even less than the distributions of $S_2^2$. The curve for Mauna Kea shows again a lower scintillation. Other observatories, except the SPM, have approximately the same characteristics. The scintillation parameters measured at SPM are greater than the typical values by about 20\%. However, we emphasize again that these results reflect each particular set of data without taking into account the seasonal distribution of the observations. It will be demonstrated in Sec.~\ref{sec:seasonal} that the $S_3^2$ parameter is subject to seasonal variations.

The differential distributions of the $S_3^2$ parameter are very similar to the differential distributions for the $S_2^2$; they have roughly the same asymmetry and a somewhat larger excess. Just as in the case of $S_2^2$, the distributions of $\ln S_3^2$ are close to normal; typical values of their skewness and excess kurtosis are $\gamma_1 \sim 0.2 - 0.4$ and $\gamma_2 \sim 0 - 0.5$.

\section{Temporal variations of the $S_2$ and $S_3$ parameters}
\label{sec:temporal}

\subsection{Nightly variations}
\label{sec:struct}

The study of temporal variations of the astroclimatic parameters on time scales of $\sim 1$~hour (in course of a night) is important in terms of short-term predictions for operational scheduling of observations \citep{Racine1996,Skidmore2009}. Of course, the variability of the parameters $S_2$ and $S_3$ is not as critical as the variability of the seeing. However, for a long time series of photometric measurements, control of the scintillation noise is also highly desirable.

To quantify the nature of the short-term variability, different approaches are used. These are the calculation of the auto-correlation function (ACF) in the usual sense \citep{Tokovinin2003MN} and the average relative or absolute differences of a parameter for a certain time delay. The fractional difference (FD) proposed in \citet{Racine1996} is not very amenable to a rigorous mathematical analysis (being a ratio, it has a non-trivial distribution function). The absolute difference (AD) used in \citet{Skidmore2009,Garcia2010} is restrictedly suitable for non-stationary processes where the variance depends strongly on the average.

To achieve the objective outlined by \citet{Racine1996} in his introduction of the FD, one could analyze logarithms of the parameters instead of the parameters themselves. This method is recommended by \citet{JenkinsWatts} for non-stationary processes of multiplicative nature, where the dispersion is nearly a linear function of the mean. We have already used such a transformation when evaluating random errors of $S_2^2$ and $S_3^2$. In Sect.~\ref{sec:results_s3} we pointed out that $\ln S_2^2$ and $\ln S_3^2$ are distributed almost normally.

Thus, to describe the temporal behaviour of the scintillation parameters we use the structure functions of their
logarithms. For the $S_2^2$ parameter: 
\begin{equation}
{\mathcal D}_2(T) = \langle (\ln S_2^2(t) - \ln S_2^2(t+T))^2\rangle_t,
\label{eq:strlog}
\end{equation}
where the averaging is performed over all available pairs for the given delay (lag) $T$. Similarly, we define the function ${\mathcal  D}_3(T)$ for the parameter $S_3^2$. In order not to complicate the analysis with normalization, we consider non-normalized functions and define the characteristic correlation time of the scintillation power using an absolute criterion.

\begin{figure}
\centering \psfig{figure=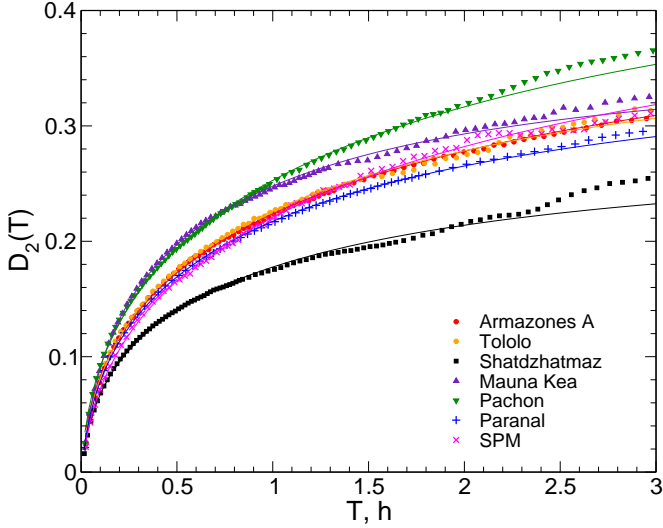,height=9cm,angle=-90}
\caption{Structure function ${\mathcal D}_2(T)$ of logarithm of the scintillation parameter $S_2^2$ at seven sites. Thin solid lines depict the approximations by Eq.~\ref{eq:stru1}.
\label{fig:s2-appr}}
\end{figure}

\begin{figure}
\centering \psfig{figure=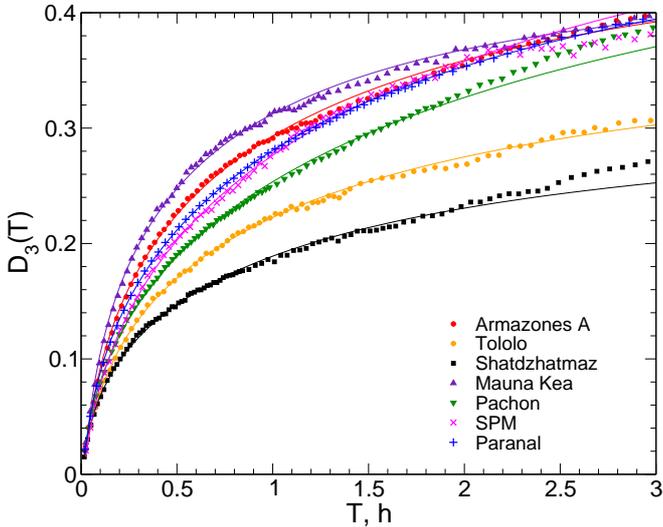,height=9cm,angle=-90}
\caption{Structure function ${\mathcal D}_3(T)$ of logarithm of the scintillation parameter $S_3^2$ at seven sites. Thin solid lines indicate the approximations by Eq.~\ref{eq:stru1}. \label{fig:s3-appr}}
\end{figure}

The structure functions for all sites are presented in Fig.~\ref{fig:s2-appr} and Fig.~\ref{fig:s3-appr}, where the contribution of uncorrelated noise was subtracted. For this calculation we used 1 minute measurements of both $S_2^2$ and $S_3^2$. Note that all structure functions have a similar behaviour for delays less than 2 hours. At longer delays, the time coverage of the individual data sets influences the computed structure functions and their behaviour becomes irregular. The initial part of the functions ${\mathcal D}_2(T)$ and ${\mathcal D}_3(T)$ is well approximated by the formula
\begin{equation}
{\mathcal D}(T) \approx \sigma^2\,[1-\exp(-\sqrt{T/\tau_c}+\delta)].
\label{eq:stru1}
\end{equation}

The main features of this dependence are infinite derivative at zero and saturation at the level of $\sigma^2$ when $T \to \infty$. The point $T = \tau_c$ corresponds to $0.632\, {\mathcal D}(\infty) = 0.632\,\sigma^2$. The time constant $\tau_c$ therefore characterizes relative changes in the power of the scintillation. The structure function (\ref{eq:stru1}) corresponds to the temporal spectral density proportional to $\propto 1/\omega$ (flicker noise) which saturates at low frequencies $\omega \ll 1/ \pi \tau_c$.

The approximation of the measured structure functions by the Eq.~(\ref{eq:stru1}) on the initial segment $T < 2$~hours shows that the parameters $\sigma^2$ and $\tau_c$ are similar at all sites, with typical values $\tau_c \sim 1$~hour. (Except for the Tolonchar site, where the time constant is longer than 10 hours). The values of $\tau_c$ are given in Table~\ref{tab:night2} and~\ref{tab:night3}. The parameter $\sigma^2$ lies in the range of $0.25 - 0.4$. The small additional parameter $\delta \sim 0.05$ compensates for the imperfect subtraction of uncorrelated noise, it is not considered here.

\begin{figure}
\centering \psfig{figure=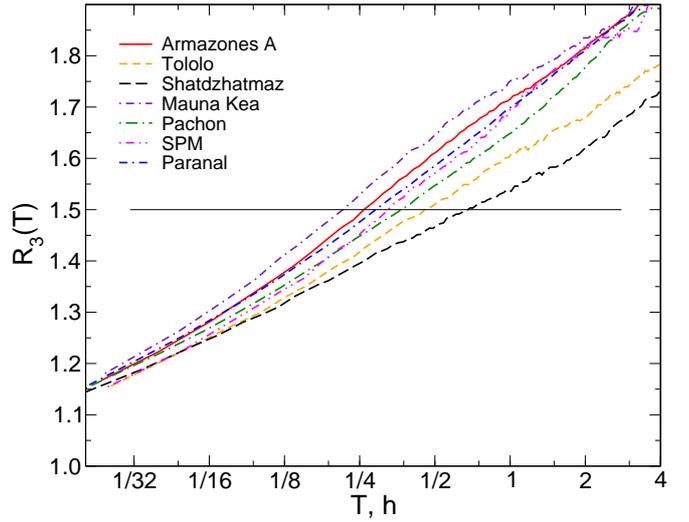,height=9cm,angle=-90}
\caption{Ratio of $S_3^2$ parameters for a time lag $T$, $R_3(T)$, at seven sites. The ratio of 1.5 times is shown by the horizontal line. \label{fig:s3-pred}}
\end{figure}

For practical purpose, another way to evaluate the stability of the scintillation noise may be more convenient. To do this, we transform the difference of the logarithms in (\ref{eq:strlog}) to a logarithm of ratio and define the statistics
\begin{equation}
R_2(T) = \exp( {\mathcal D}_2(T)^{1/2} ) ,
\end{equation}
and the similar function $R_3(T)$ for the $S_3^2$ parameter. The functions $R_3(T)$ are plotted in  Fig.~\ref{fig:s3-pred}. The time lags $\tau_{1.5}$ corresponding to changes of $S_2^2$ and $S_3^2$ by 1.5 times are given in Table~\ref{tab:night2} and~\ref{tab:night3}.

At almost all sites, the $S_3^2$ parameter is changing faster than the $S_2^2$. Apparently, in this case the variations of the wind speed are important. The fastest changes of the scintillation power are observed at Mauna Kea: $0.2$ and $0.3$ hours. The atmosphere above Mt.~Shatdzhatmaz is changing most slowly: $0.7$ and $0.8$ hours. However, note that most measurements at this site have been performed in the fall which is the most stable season.

Variation of the scintillation power over longer time scales can be estimated by comparing its variability within a night to the variability during the measurement season. To do this for both scintillation parameters, we calculated the average $\langle \ln S^2 \rangle_j$ and the corresponding variances $\sigma^2_j$ within each night $j$. Then, the variance of the average nightly quantities $\sigma^2_m$ was evaluated. The intra-night variability is quantified by its mean over the whole campaign, $\langle \sigma^2_j \rangle$. Nights containing less than 10 measurements of the $S_2^2$ and $S_3^2$ were discarded in this analysis.

The $\langle \sigma^2_j \rangle$, corrected for the contribution of random measurement errors, are presented in Table~\ref{tab:night2} and~\ref{tab:night3}. On average, the variance of the random noise is $0.01$ for the $S_2^2$ parameter, and about $0.1$ for the $S_3^2$ (see Table~\ref{tab:s2} and~\ref{tab:s3}). It is seen that the relative variability within a night is almost identical at all observatories. As expected, the variability of the $\ln S_3^2$ is slightly higher than the variability of the $\ln S_2^2$. There is one statistically insignificant exception for the measurements carried out at Tololo. The average value of the variability within nights for $\ln S_2^2$ is $0.13\pm 0.01$, for $\ln S_3^2$ is $0.16 \pm 0.03$.

The variability of the nightly averages is a combination of regular seasonal changes (see the next Section) and the sporadic variability from night to night. We did not isolate the seasonal behaviour, so the full variances $\sigma^2_m$ are listed in Table~\ref{tab:night2} and~\ref{tab:night3}. Not surprisingly, the maximum values are observed for Shatdzhatmaz and S.~Pedro Martir, as the seasonal variability at these sites is as large as $0.1 - 0.14$ (about 2.5 times larger than the intra-night variability). Measurements at Tolonchar also show a significant long-term variability. Many parameters describing the scintillation noise at this site are substantially different from the other sites. The median variability from night to night at Tolonchar is $0.31 \pm 0.07$ for $\ln S_2^2$ and $0.24 \pm 0.08$ for $\ln S_3^2$ (less than for $\ln S_2^2$, which is atypical).

\begin{table}
\caption{ Variation of $\ln S_2^2$ during the night and between the nights. The time constants $\tau_c$ and $\tau_{1.5}$ are given in hours.\label{tab:night2}}
\centering
\begin{tabular}{lrrrrrrr}
\hline\hline
Site & $N_n$ & $\sigma^2_0 $ & $\sigma^2_m$ & $\langle\sigma^2_j\rangle$ & $\tau_c$ & $\tau_{1.5}$ \\
\hline
Armazones A & 928& 0.399& 0.267& 0.130 & 1.53 & 0.43\\
Armazones B & 525& 0.459& 0.310& 0.132 & & \\
La Chira & 201& 0.411& 0.271& 0.124 & 2.46 & 0.36 \\
Mauna Kea & 576& 0.498& 0.361& 0.141 & 0.58 & 0.31 \\
Pach\'on &1426& 0.469& 0.339& 0.144 & 1.79 & 0.33 \\
Paranal A &1365& 0.416& 0.293& 0.123 & 1.01 & 0.46\\
Paranal B & 582& 0.406& 0.281& 0.125 & 1.01 & 0.46 \\
S.~Pedro Martir& 608& 0.519& 0.378& 0.126 & 2.08 & 0.48 \\
Shatdzhatmaz & 706& 0.472& 0.417& 0.094 & 0.84 & 0.80\\
Tolar & 264& 0.355& 0.209& 0.119 & 1.04 & 0.55 \\
Tololo & 625& 0.427& 0.328& 0.121 & 1.13 & 0.42 \\
Tolonchar & 386& 0.575& 0.477& 0.135 & $>$10& 0.44 \\
Ventarrones & 553& 0.439& 0.299& 0.124 & 0.99 & 0.47 \\
\hline\hline
\end{tabular}
\end{table}

\begin{table}
\caption{ Variation of $\ln S_3^2$ during the night and between the nights. The time constants $\tau_c$ and $\tau_{1.5}$ are given in hours. \label{tab:night3}}
\centering
\begin{tabular}{lrrrrrrr}
\hline\hline
Site & $N_n$ & $\sigma^2_0 $ & $\sigma^2_m$ & $\langle\sigma^2_j\rangle$ & $\tau_c$ & $\tau_{1.5}$ \\
\hline
Armazones A & 929& 0.487& 0.206& 0.173 & 0.82 & 0.26 \\
Armazones B & 526& 0.559& 0.244& 0.215 & & \\
La Chira & 202& 0.495& 0.240& 0.191 & 0.82 & 0.18 \\
Mauna Kea & 577& 0.501& 0.232& 0.169 & 0.48 & 0.21 \\
Pach\'on &1770& 0.488& 0.243& 0.149 & 2.36 & 0.37 \\
Paranal A &1371& 0.465& 0.229& 0.156 & 1.43 & 0.29 \\
Paranal B & 583& 0.457& 0.216& 0.179 & 1.43 & 0.29 \\
S.~Pedro Martir& 609& 0.570& 0.348& 0.144 & 2.00 & 0.32 \\
Shatdzhatmaz & 712& 0.621& 0.451& 0.106 & 0.97 & 0.68 \\
Tolar & 264& 0.388& 0.149& 0.116 & 0.53 & 0.33 \\
Tololo & 630& 0.460& 0.258& 0.113 & 1.00 & 0.45 \\
Tolonchar & 409& 0.619& 0.342& 0.176 & $>$10& 0.27 \\
Ventarrones & 563& 0.486& 0.234& 0.154 & 0.69 & 0.27 \\
\hline\hline
\end{tabular}
\end{table}

\subsection{Seasonal variability}
\label{sec:seasonal}

\begin{figure*}
\tabcolsep=0pt
\begin{tabular}{ccc}
\psfig{figure=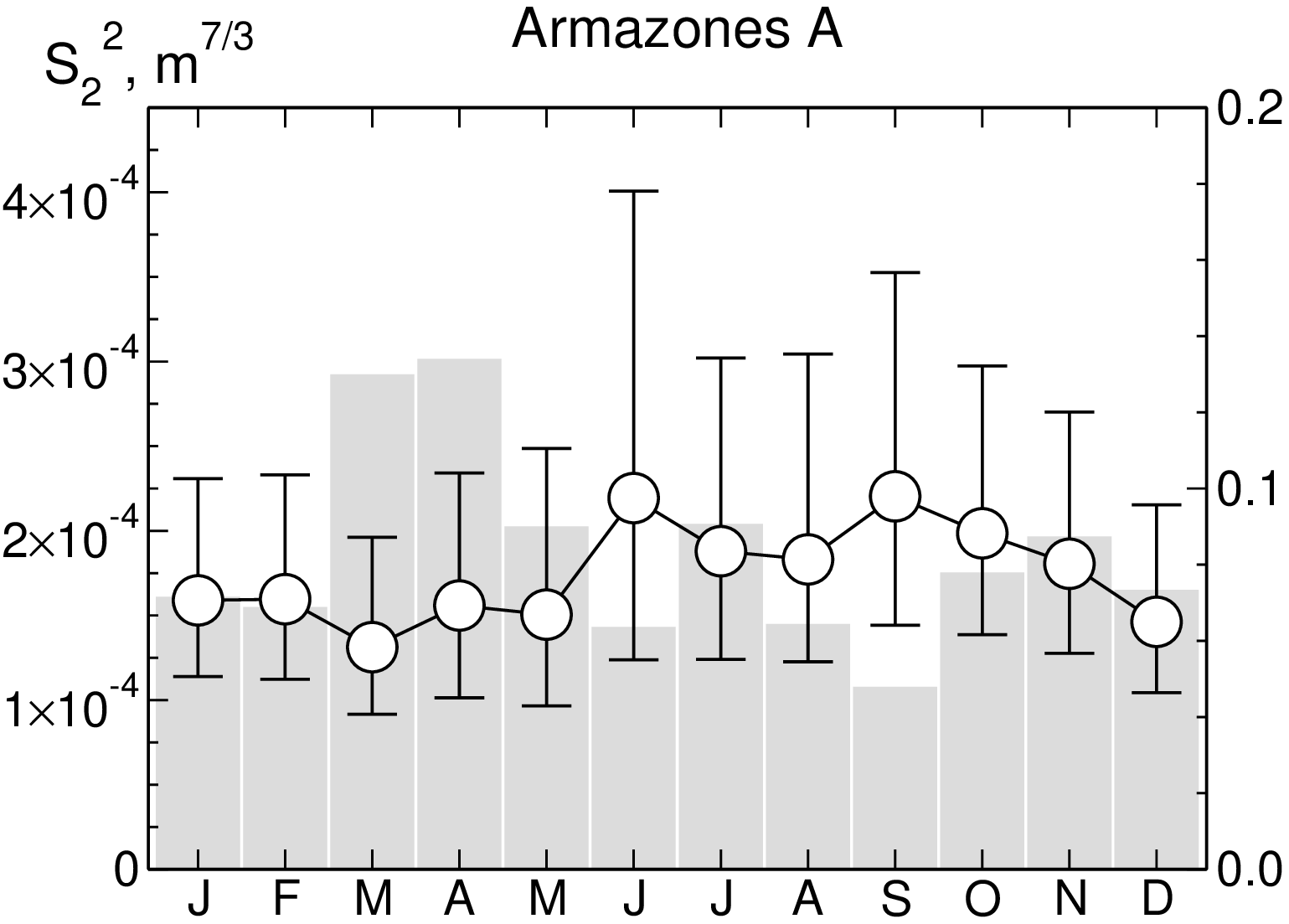,height=6.15cm,angle=-90} &            
\psfig{figure=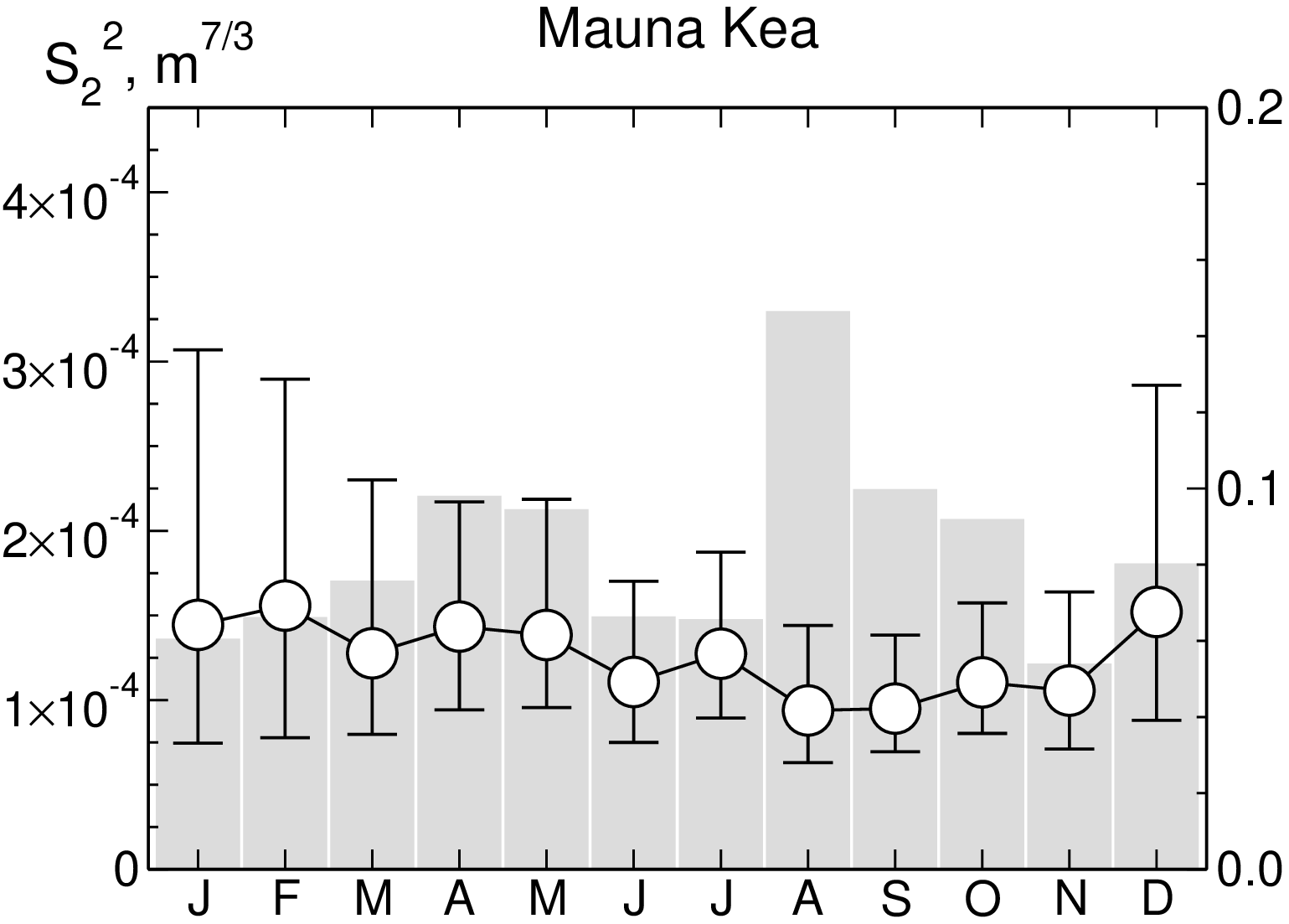,height=6.15cm,angle=-90} &            
\psfig{figure=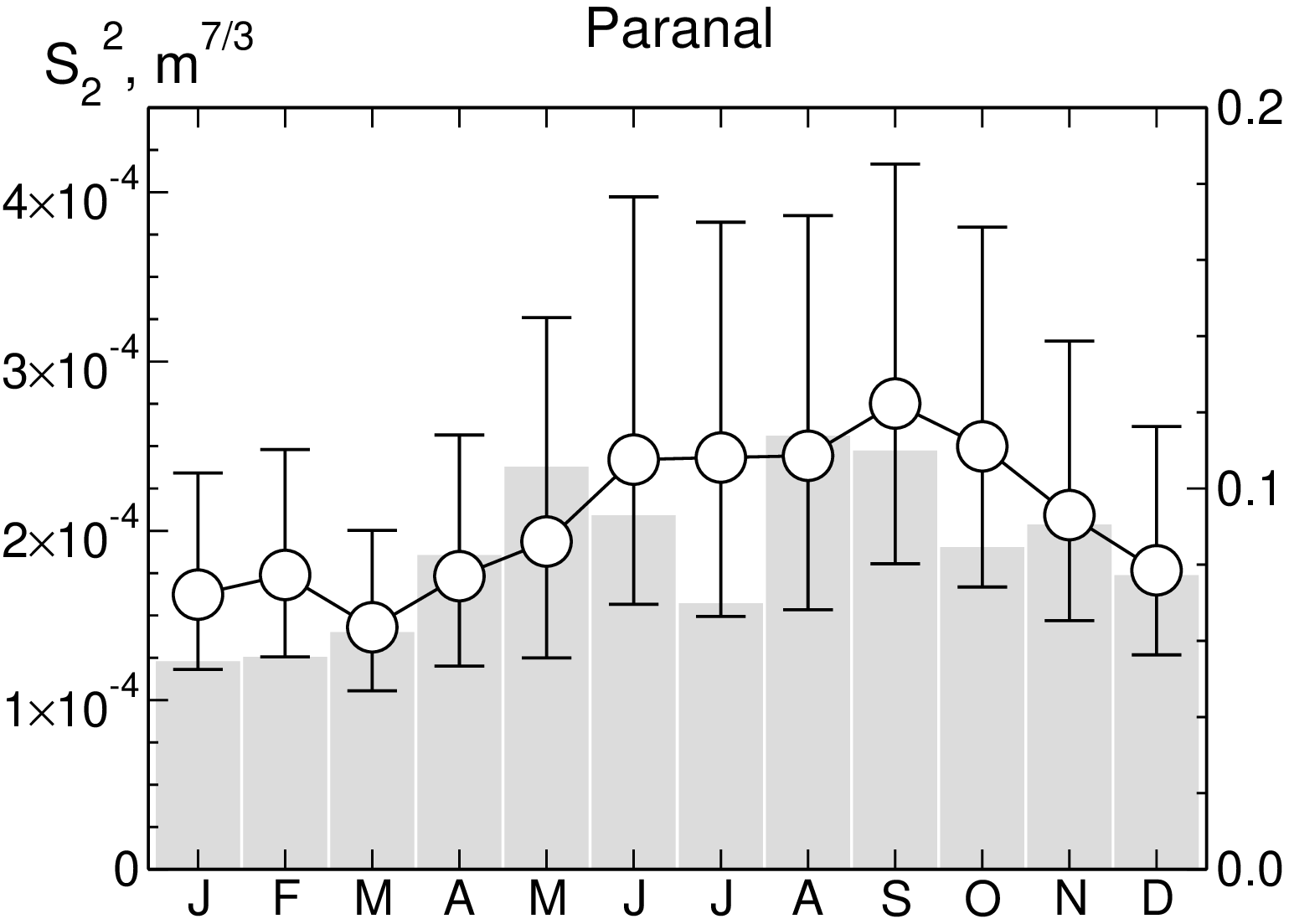,height=6.15cm,angle=-90} \\[-12pt]    
\psfig{figure=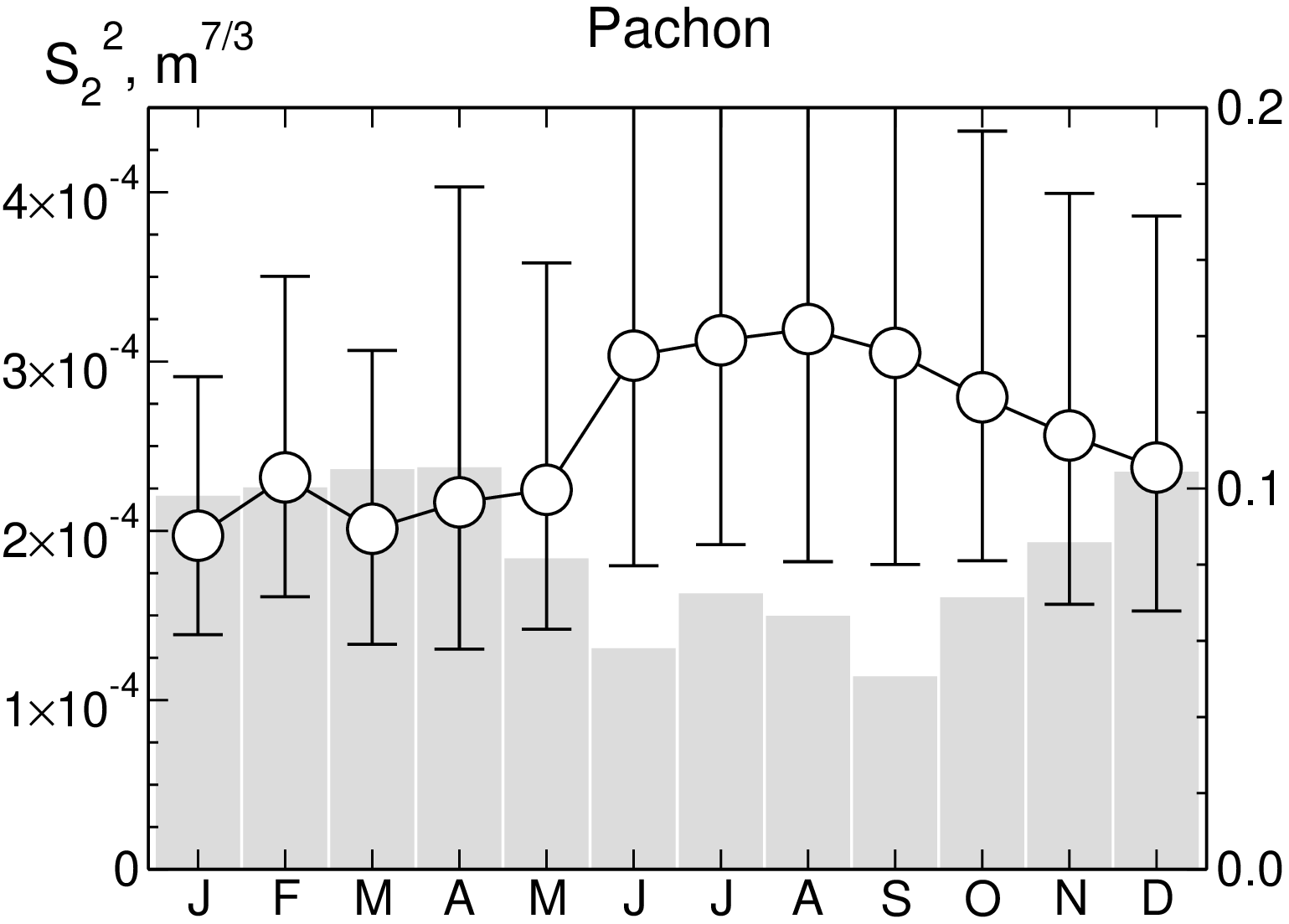,height=6.15cm,angle=-90} &            
\psfig{figure=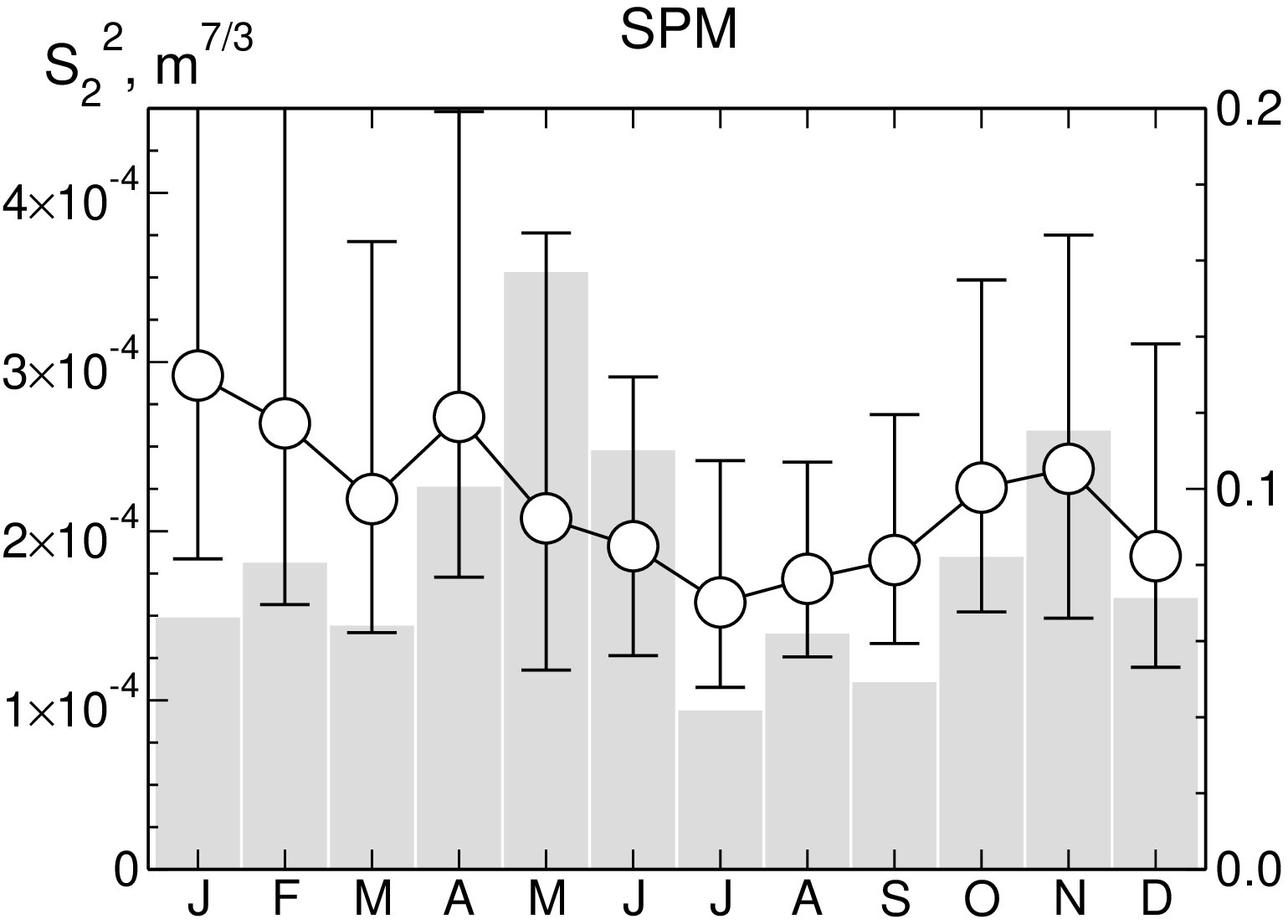,height=6.15cm,angle=-90} &            
\psfig{figure=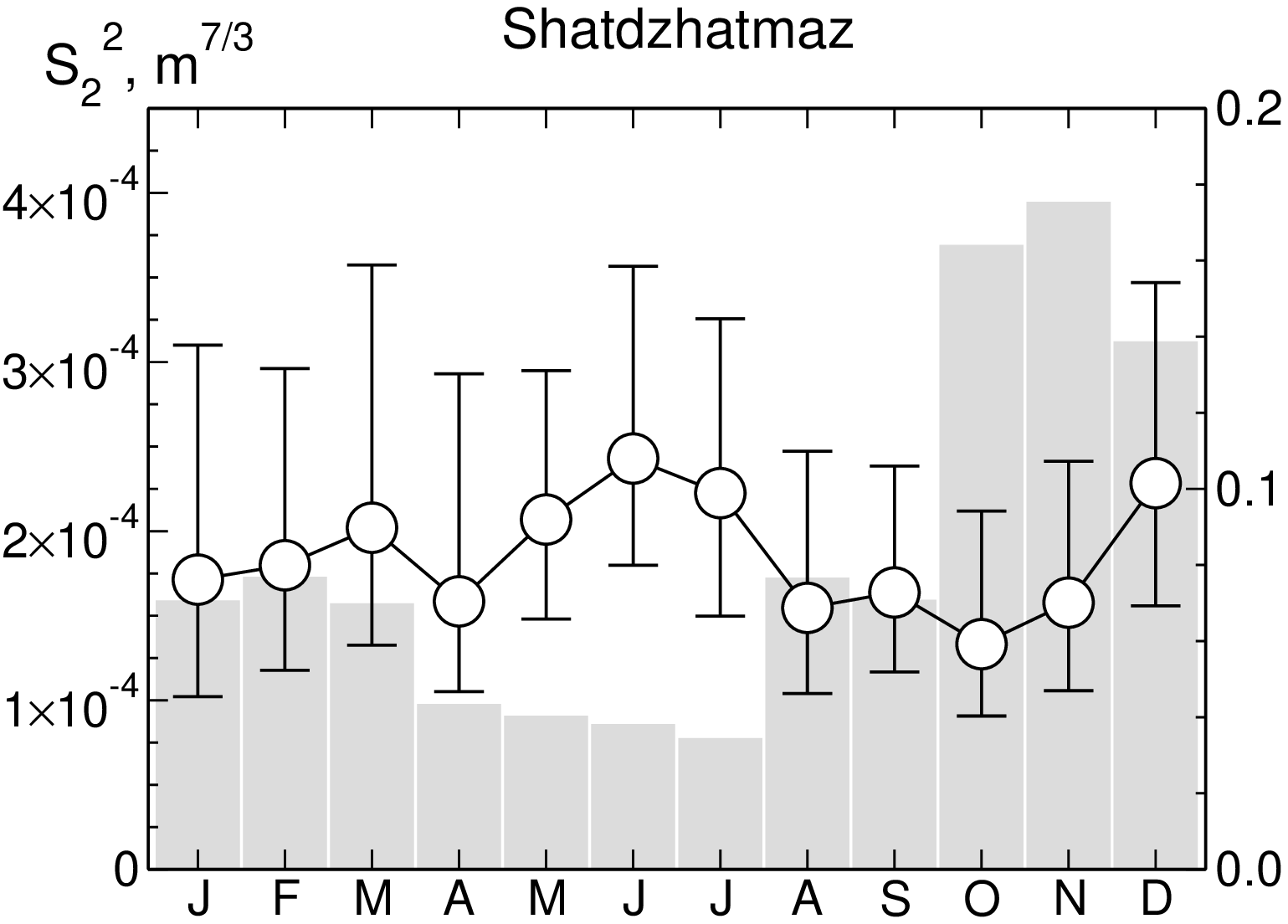,height=6.15cm,angle=-90} \\[-12pt]    
\psfig{figure=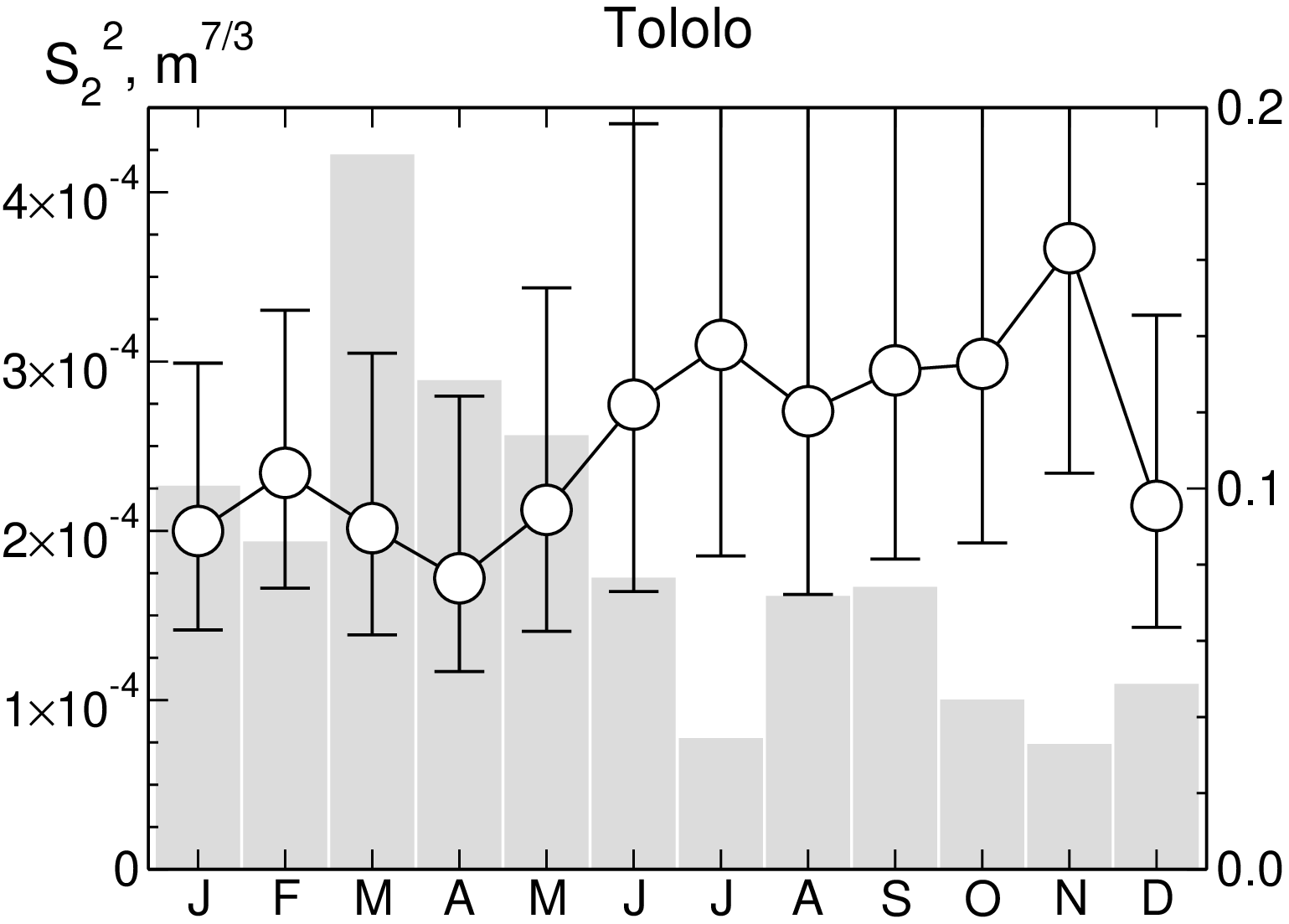,height=6.15cm,angle=-90} &            
\psfig{figure=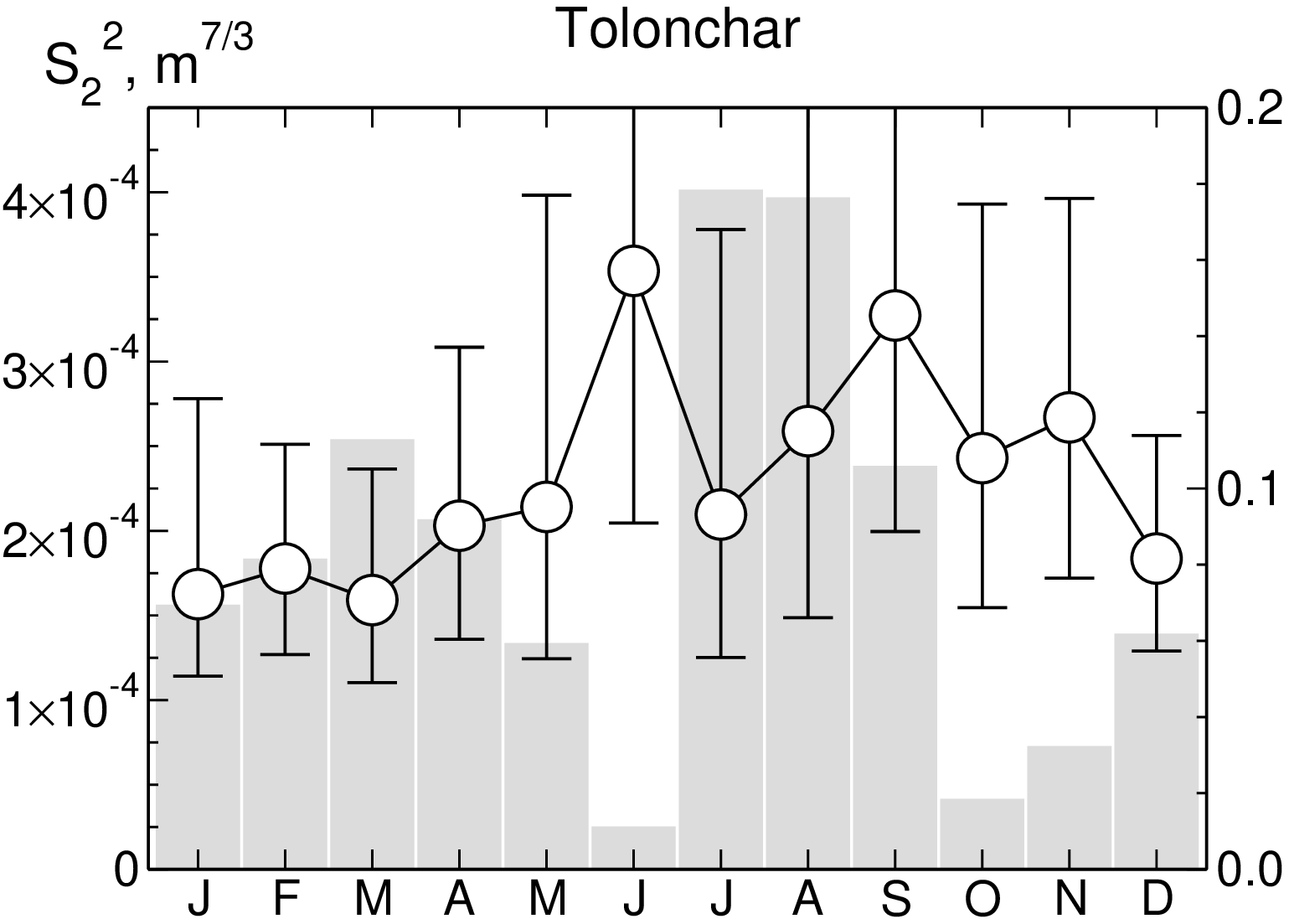,height=6.15cm,angle=-90} &            
\psfig{figure=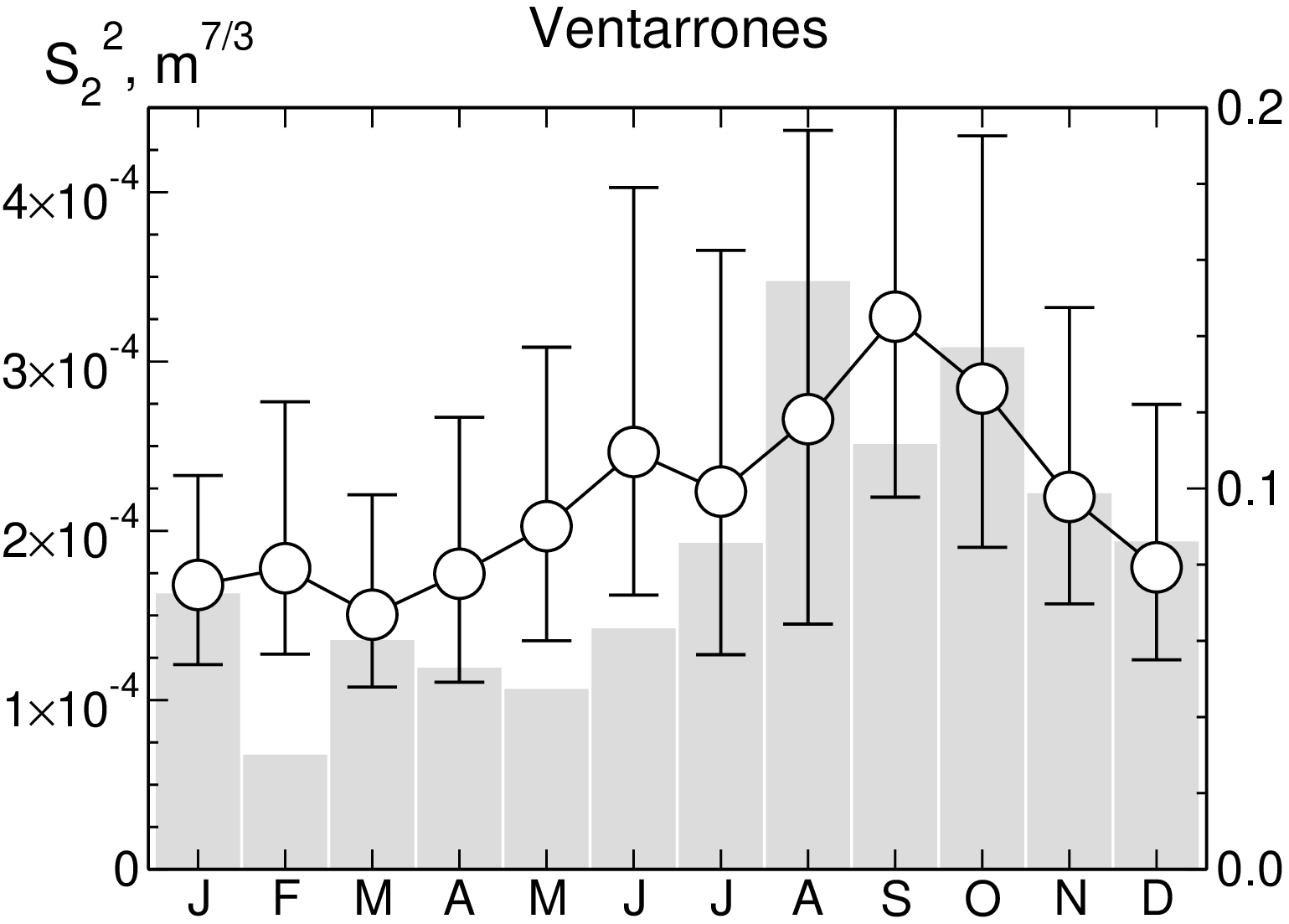,height=6.15cm,angle=-90} \\[-12pt]    
\end{tabular}
\caption{Seasonal variation of the scintillation parameter $S_2^2$. Monthly medians are plotted as circles, the quartiles are indicated by bars. The grey bars show {\bf the share of the measurements (right axis)} entering the respective month.
\label{fig:s2-seas}}
\end{figure*}

\begin{figure*}
\tabcolsep=0pt
\begin{tabular}{ccc}
\psfig{figure=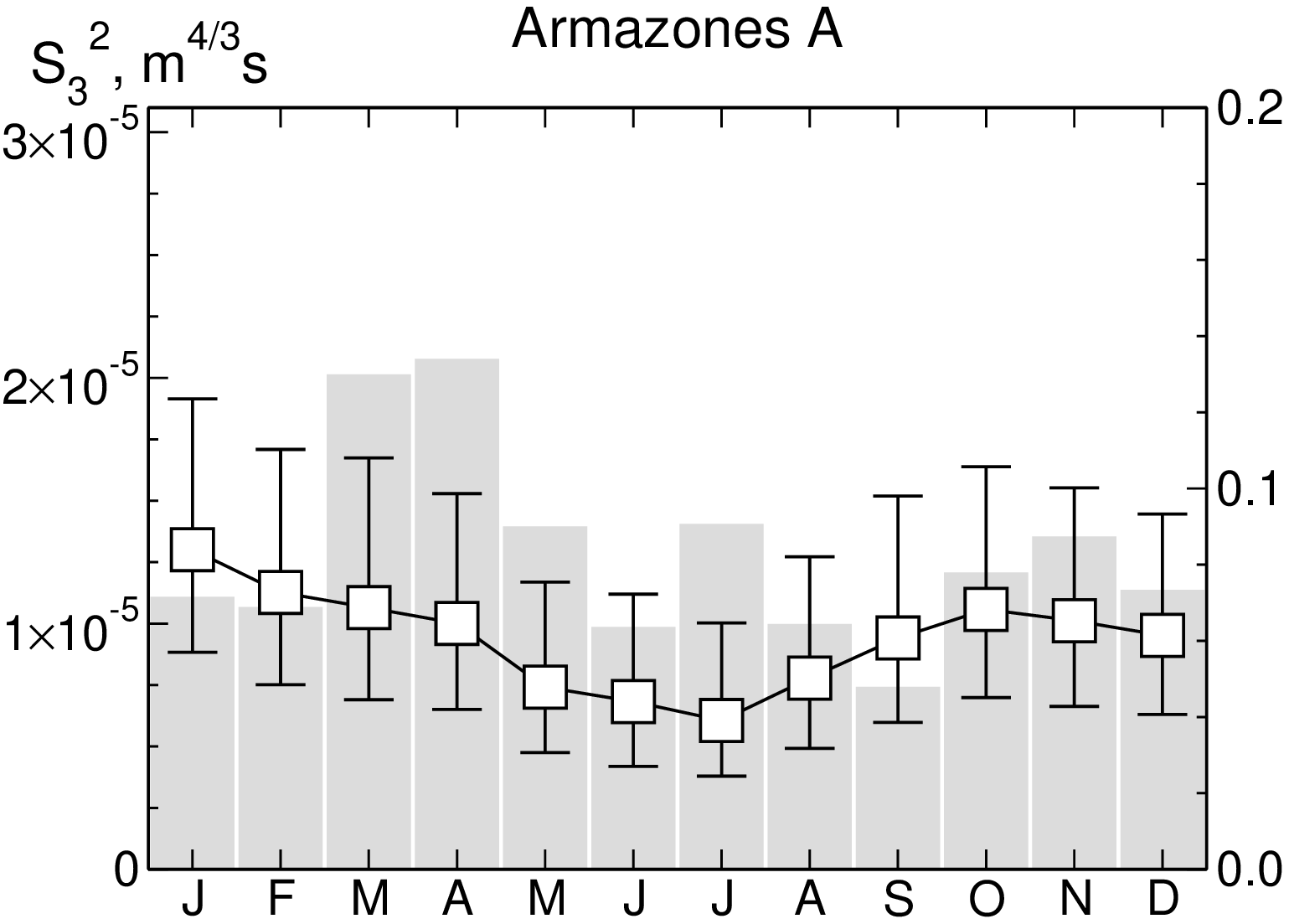,height=6.15cm,angle=-90} &            
\psfig{figure=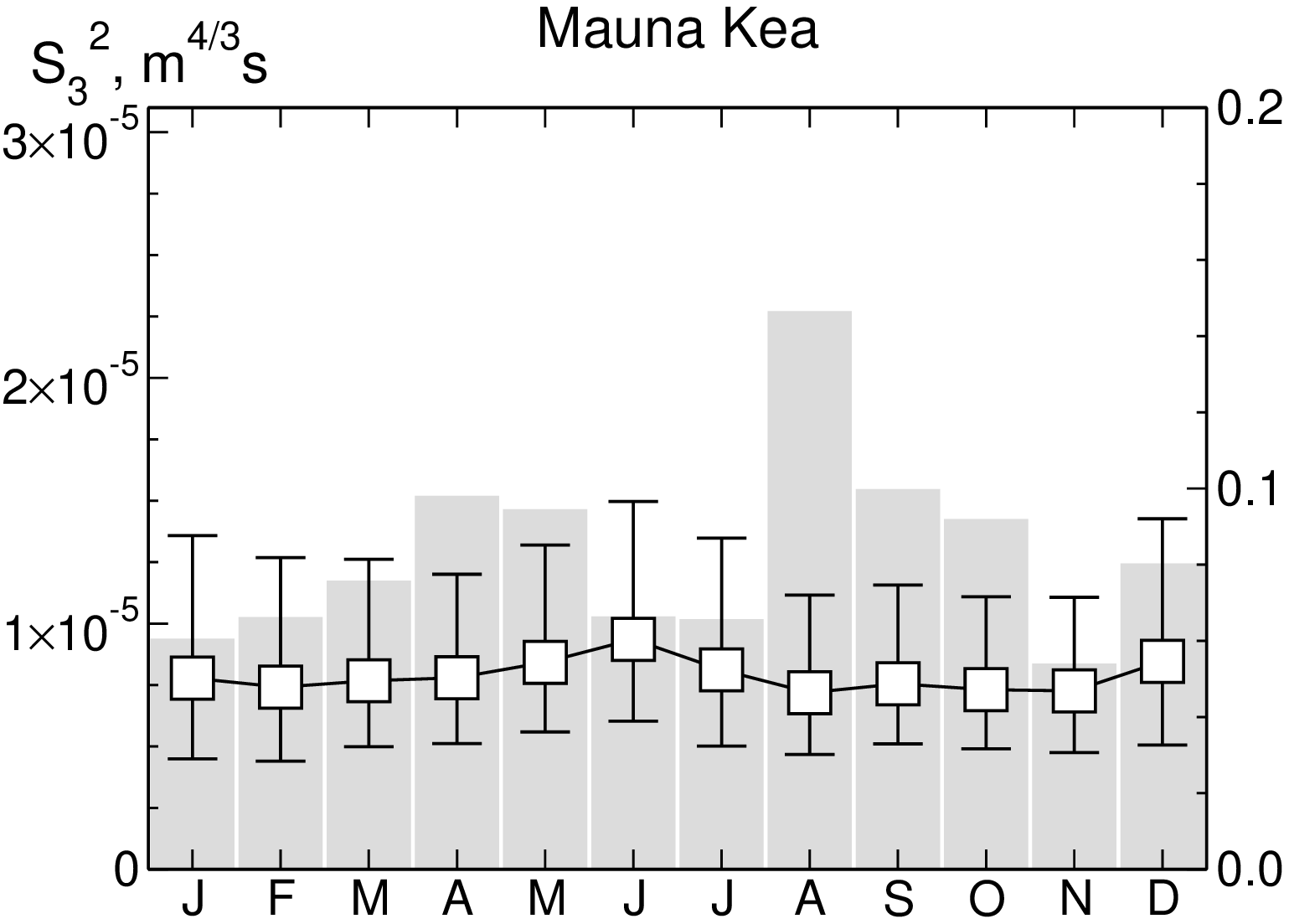,height=6.15cm,angle=-90} &            
\psfig{figure=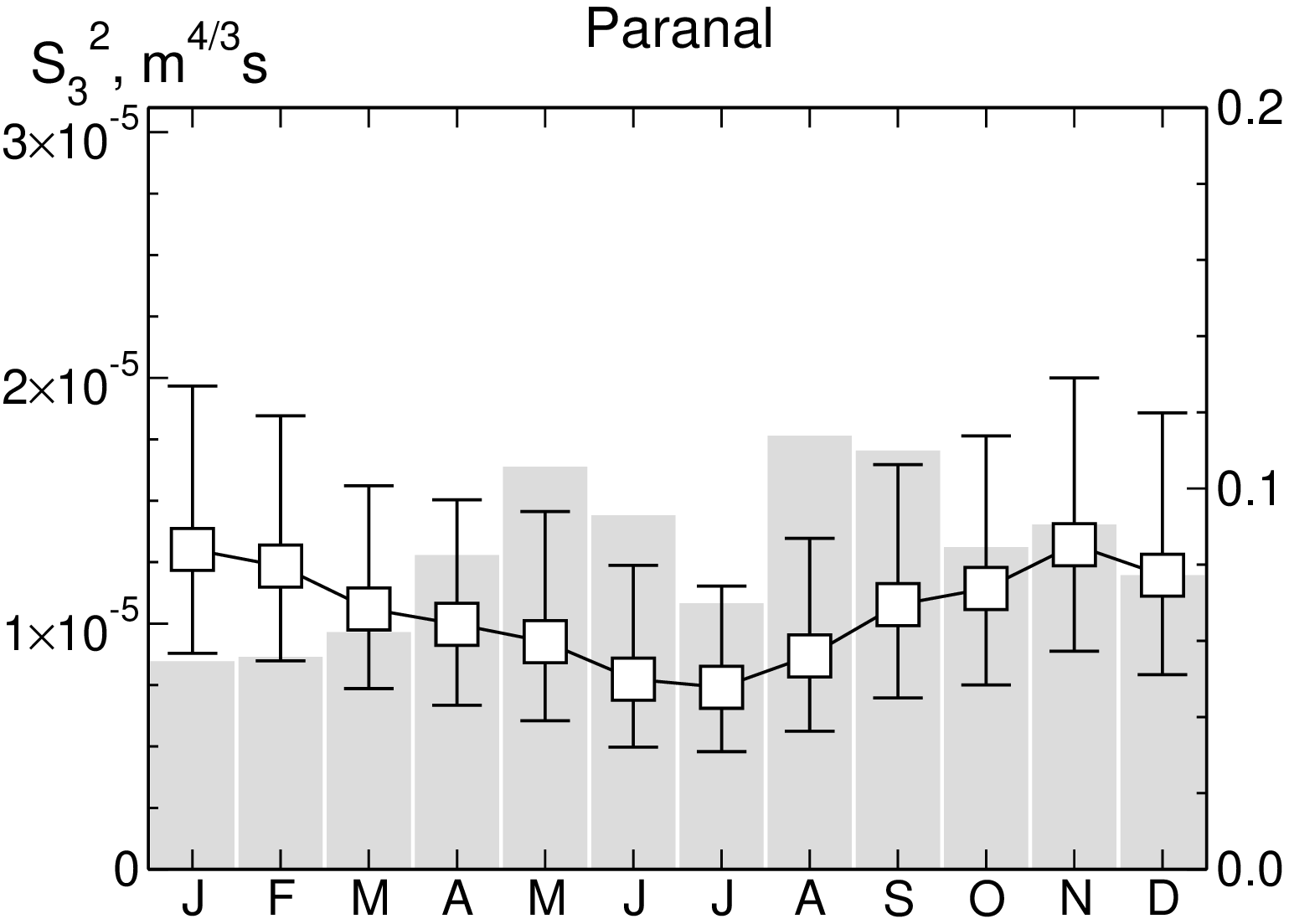,height=6.15cm,angle=-90} \\[-12pt]    
\psfig{figure=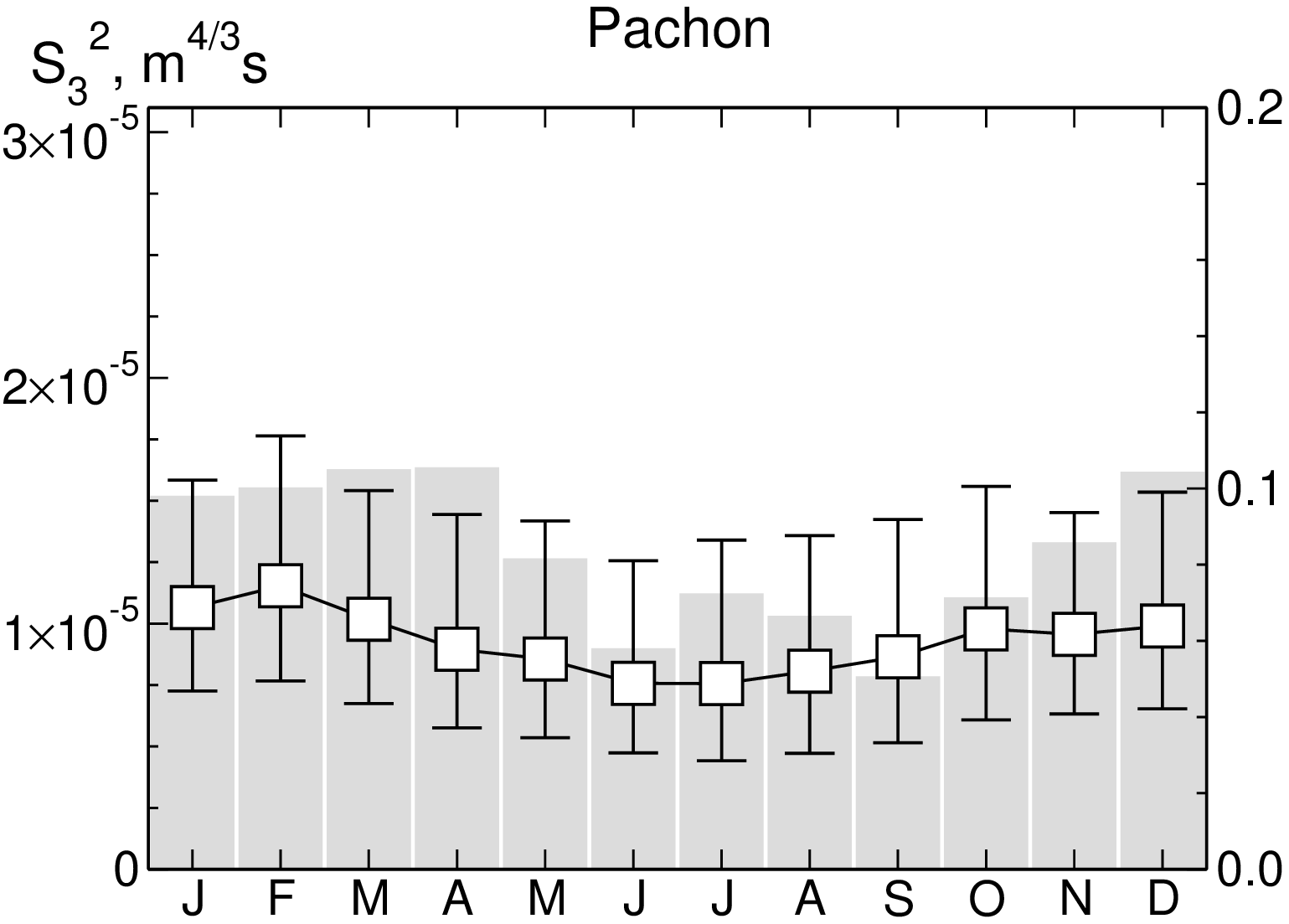,height=6.15cm,angle=-90} &            
\psfig{figure=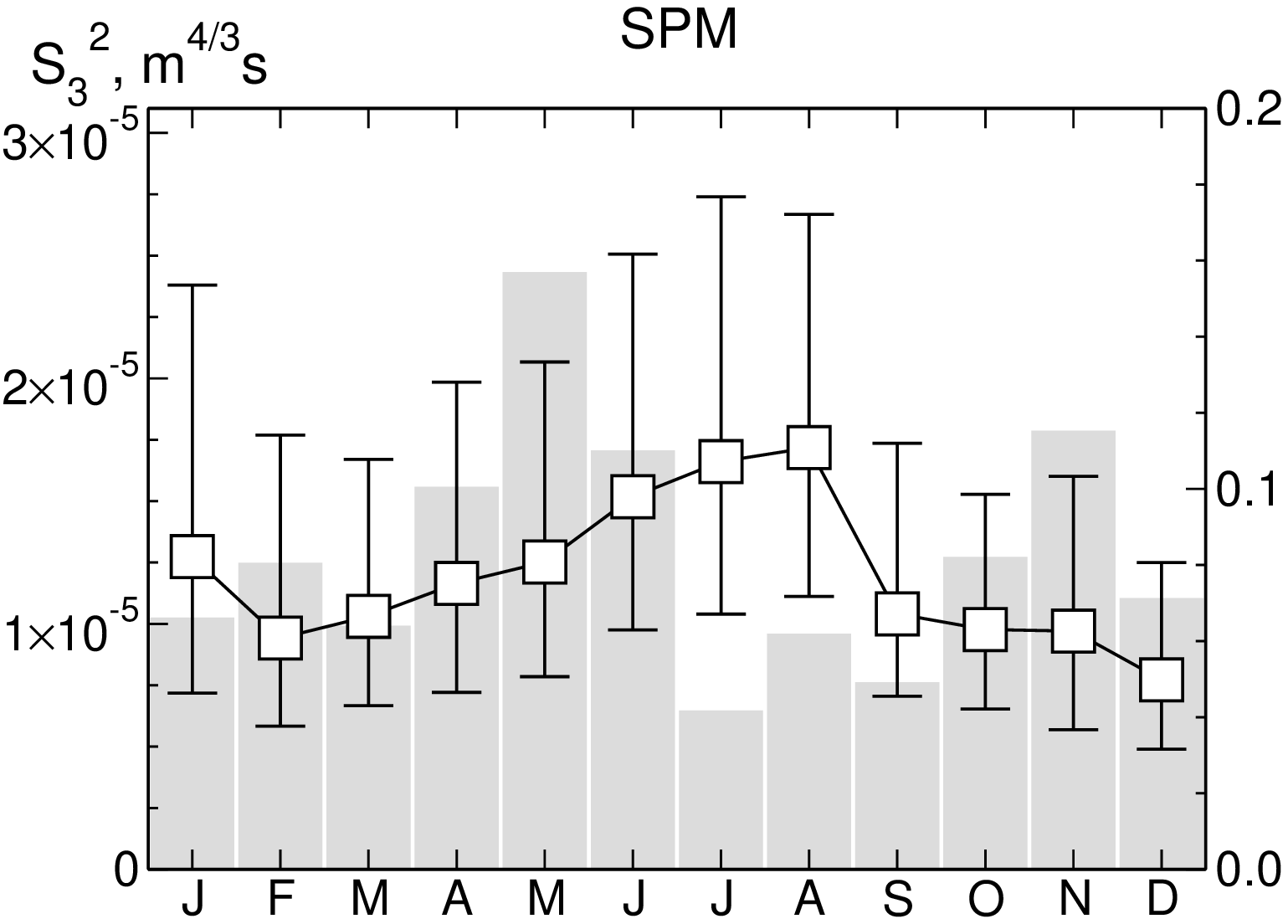,height=6.15cm,angle=-90} &            
\psfig{figure=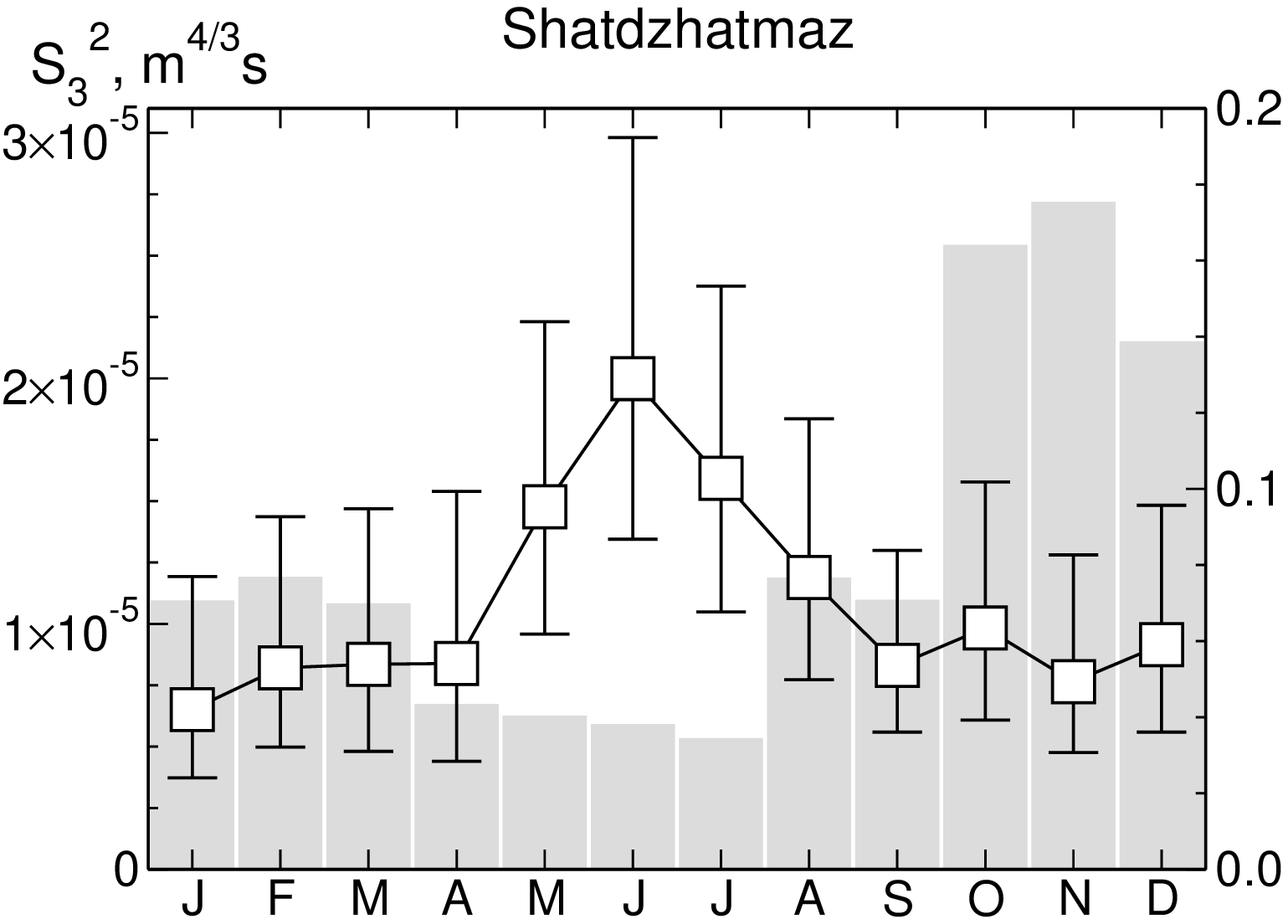,height=6.15cm,angle=-90} \\[-12pt]    
\psfig{figure=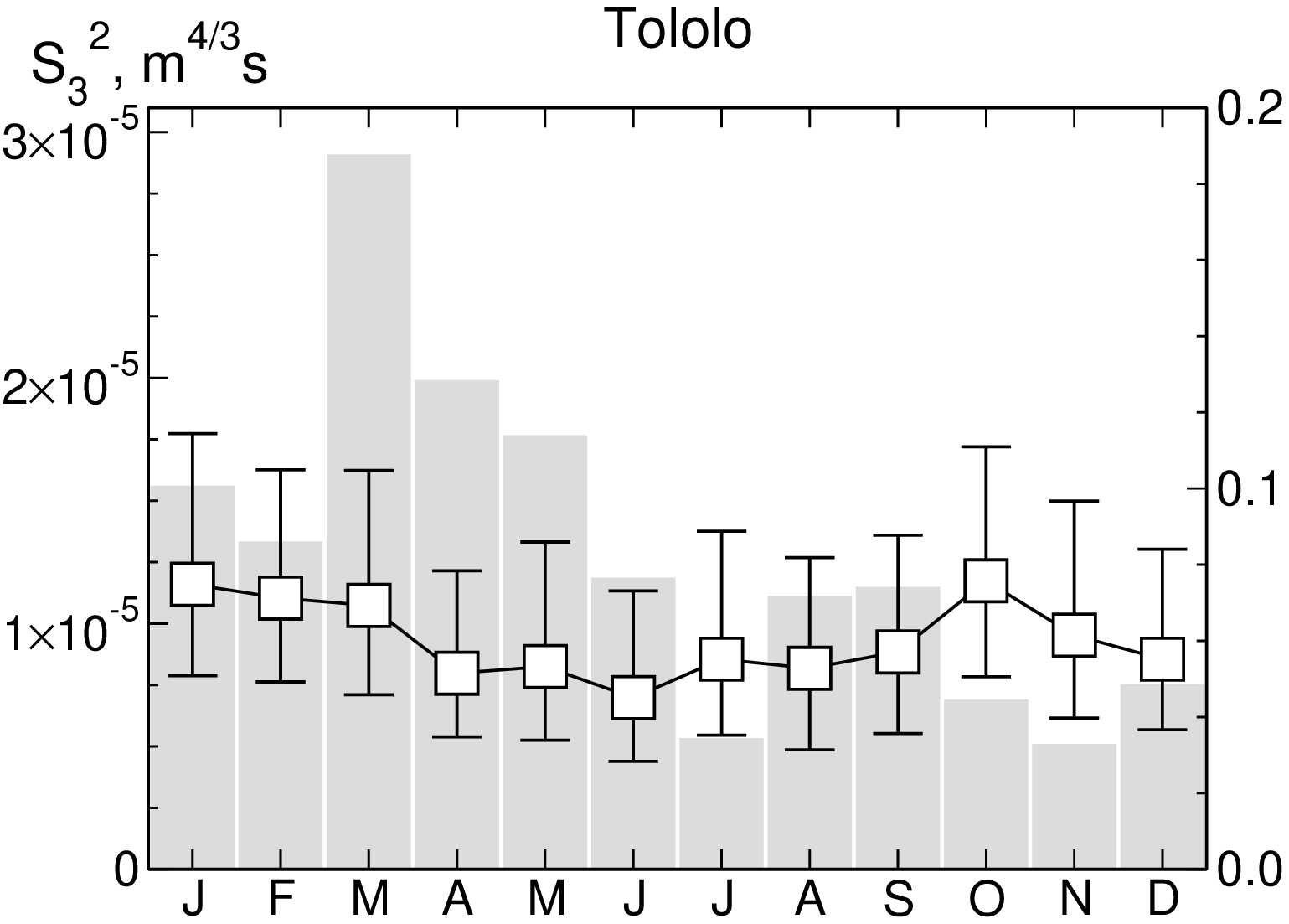,height=6.15cm,angle=-90} &            
\psfig{figure=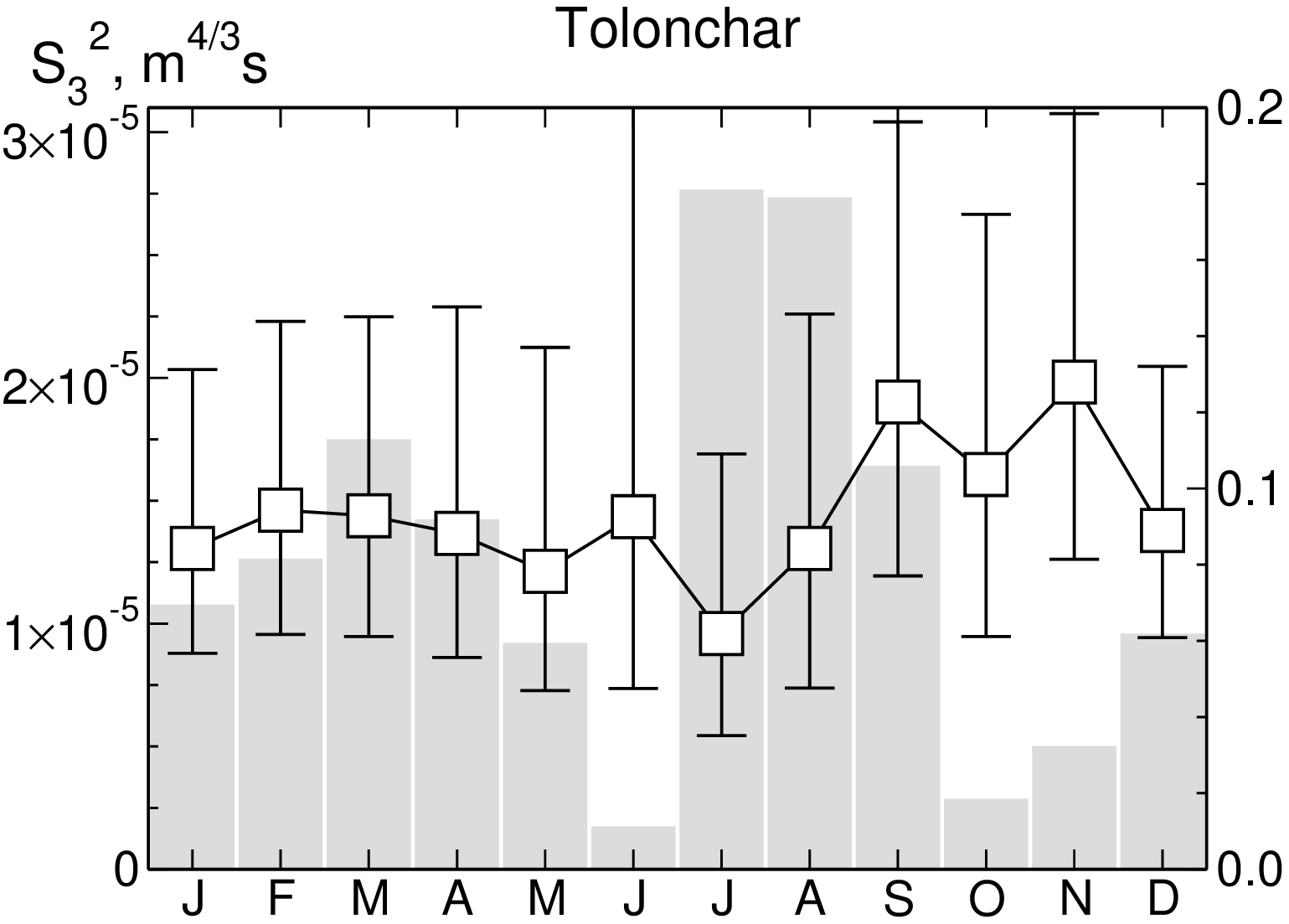,height=6.15cm,angle=-90} &            
\psfig{figure=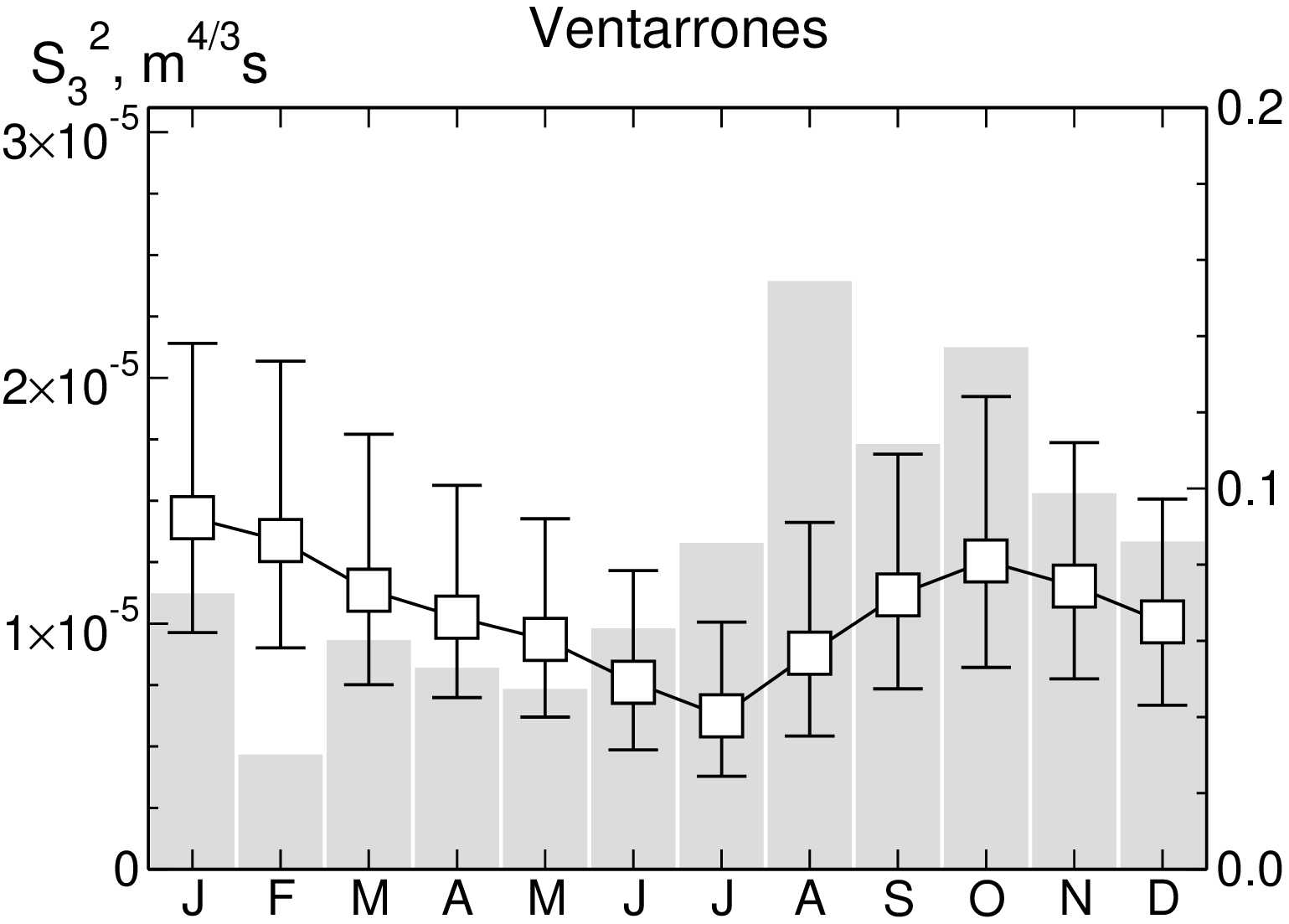,height=6.15cm,angle=-90} \\[-12pt]    
\end{tabular}
\caption{Seasonal variation of the scintillation parameter $S_3^2$. Monthly medians are plotted as circles, the quartiles are indicated by bars. The grey bars show {\bf the share of the measurements (right axis)} entering the respective month. \label{fig:s3-seas}}
\end{figure*}

For the all observatories, the long-term evolution of the parameters $S_2^2$ and $S_3^2$ is seasonal. Seasonal variability is evident in the raw data, but for a more detailed and quantitative study, we calculated the statistical characteristics by months of the year. These medians and quartiles (in the form of bars) for the parameter $S_2^2$ are shown in Fig.~\ref{fig:s2-seas} and for $S_3^2$ in Fig.~\ref{fig:s3-seas}.

Note that almost all the samples on which these statistics were determined are sufficiently large and their volumes vary by no more than 2 times from month to month. The exception is the Mt.~Shatdzhatmaz data, where the vast majority of clear weather happens during the fall. The seasonal dependence of $S_3^2$ for Armazones~B (not plotted) is similar to that of Armazones~A, with slightly larger values.

In these plots, the noticeable difference between the northern and southern hemispheres is seen, especially for $S_3^2$. There is also some dependence of the seasonal variability on the latitude of the observatory. The lowest seasonal variability is found at Mauna Kea (for both $S_2^2$ and $S_3^2$). The most pronounced seasonal variability of $S_3^2$ can be seen on the curve for Shatdzhatmaz in the summer months; the scintillation noise increased by almost two times, although this makes little effect on the overall median (see Table~\ref{tab:s3}). In contrast, the data for the SPM observatory contain a large number of measurements from May to August, which over-estimates the median for the entire data set.

Conversely, in the southern hemisphere the median of $S_3^2$ is minimal in the period from May to September, and the median $S_2^2$ is maximal from July to September, while in the northern hemisphere there is a minimum in this season. However, in general the seasonal variability does not exceed the amplitude of distributions defined by the quartiles. This means that at any time of the year there is a reasonable probability of both good and bad conditions for high-precision photometry.

Seasonal variations in the scintillation noise are caused by the redistribution of seasonal winds in the upper atmosphere. Moreover, an increase in the wind speed leads to an increase in the $S_2^2$ parameter but also to a decrease of the $S_3^2$. Data on the wind speed above the tropopause may also be obtained from our measurements.

\subsection{Correlation between close observatories}
\label{sec:correl}

Strong correlation of scintillation power on time scales of the order of $0.5 -1$ hours (Sect.~\ref{sec:struct}) implies that the size of the spatial region in which the power of turbulence, wind, and their distribution over altitude can be considered constant, is $\sim$100~km. In such case, a strong correlation of the scintillation power should be observed for closely spaced observatories ($<$50~km).

Some observatories on our list are geographically located very close to each other. The distance between Tololo and Pach\'on is about 10\,km. The distance between Paranal and Armazones is also no more than 25\,km. Additionally, in an area around Paranal, the ESO team investigated two sites: La Chira (15 km) and Ventarrones (32 km). Two other summits in northern Chile were studied by the TMT team: Tolar and Tolonchar are spaced from Paranal by about 300 km, first to the North and the other to the West.

\begin{figure}
\centering
\psfig{figure=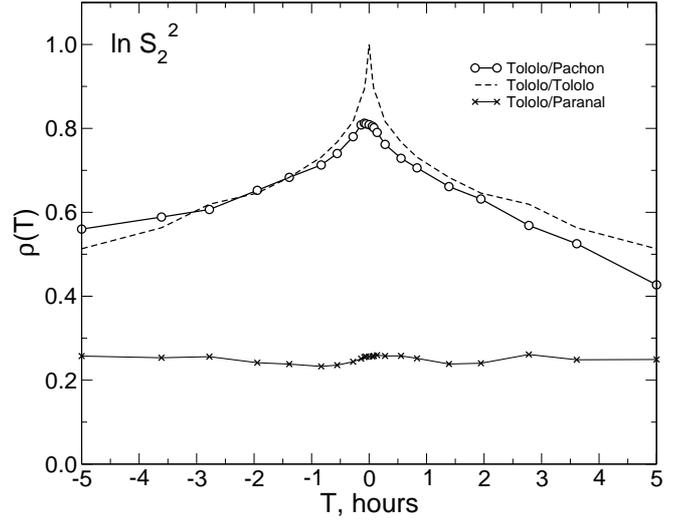,height=9cm,angle=-90}
\caption{
Cross-correlation between the $\ln S_2^2$ measurements at Tololo and Pach\'on. For comparison, the auto-correlation function for Tololo is plotted (dashed line). The cross-correlation between Tololo and Paranal (crosses) is close to zero. \label{fig:s2cross-ctio}}
\end{figure}

The cross-correlation of the scintillation parameter $S_2^2$ (or $S_3^2$) measured at two observatories, was calculated as follows. For the required time delay $T$, the pairs of measurements for which the difference between their acquisition time is less than $T/2$ (30~s) were selected from the two data sets. Then, using the classical method, the correlation coefficient $\rho(T)$ of the first sample on the second for the logarithms of the scintillation parameters $\ln S^2$ was calculated. Finally, we correct the correlation coefficient for the contribution of random measurement errors by its renormalization with corrected variances.

Naturally, the correlation reflects not only the short-term variations of the scintillation power, but also its variability on the time scale of days, including seasonal variability. This can be clearly seen in Fig.~\ref{fig:s2cross-ctio} which shows the cross-correlation between Tololo and Pach\'on with time lags $-5 < T < +5$ hours. In addition to the correlation peak of $\sim 1$ hour width caused by the intra-night variability, there is a wide pedestal at the level of $\rho(T) \approx 0.5$. The cross-correlation between measurements at Tololo and Paranal, also depicted in the figure, shows no correlation peak, while its constant level of $\rho(T) \approx 0.2$ reflects the common nature of seasonal variability at observatories in Chile. In contrast, a typical cross-correlation between Paranal and Shatdzhatmaz is $\approx -0.2$ (the seasonal variability is anti-correlated).

\begin{figure}
\centering
\psfig{figure=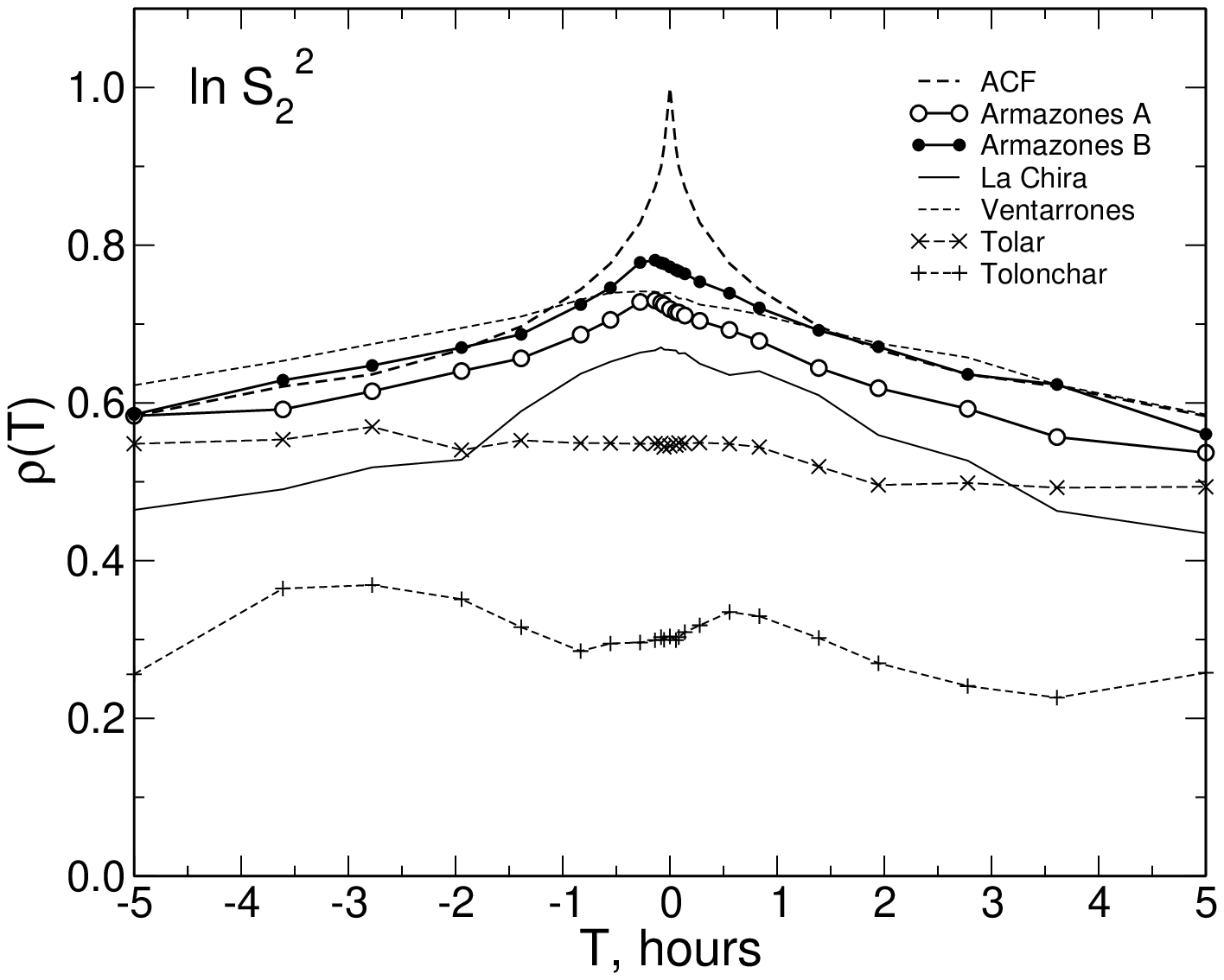,height=9cm,angle=-90}
\caption{ Same as in Fig.~\ref{fig:s2cross-ctio} for Paranal and the other nearby sites. For comparison, the auto-correlation function for Paranal is plotted in dashed line. Short-term cross-correlations between Paranal and Tolar (crosses) and Tolonchar (pluses) are absent. \label{fig:s2cross-par}}
\end{figure}

The width of the correlation peak is comparable to the width of the peak of the auto-correlation function (ACF) for each individual site. For illustration, the ACF for Tololo is plotted in Fig.~\ref{fig:s2cross-ctio}. A similar situation is observed for other close observatories, see Fig.~\ref{fig:s2cross-par}, except for weak cross-correlation peak between Paranal and Ventarrones. The maximum cross-correlation is larger for the parameter $S_2^2$ than for $S_3^2$, but not by a significant amount. The height of the peak above the extended pedestal is approximately $0.2$. Generally, when there is a significant correlation, the ratio of the cross-correlation peak to the pedestal is in good agreement with the relative variability during the night and the variability from night to night (Tables~\ref{tab:night2} and~\ref{tab:night3}).

The lack of a significant delay in the variations of the scintillation power indicates that at large spatial scales ($10 - 100$~km), the wind transportation of the OT is not dominant compared to the intrinsic evolution of the turbulence.

\section{Discussion and conclusion}

The resulting characteristics of the scintillation noise in the regime of short exposures are shown in Table~\ref{tab:comp2}, for long exposures -- in Table~\ref{tab:compar}. Instead of the squared parameters, those tables list $S_2$ and $S_3$. The $S_3$ was calculated as the average between the values obtained under the assumptions of longitudinal and transverse wind. For the Armazones site, the average values computed with a weight proportional to the number of measurements for campaigns Armazones~A and Armazones~B are presented.

Recall that the scintillation noise at the zenith $\sigma_\mathrm{S}$ and $\sigma_\mathrm{L}$ for a telescope with diameter $D$ is calculated by the formulae
\begin{equation}
\sigma_\mathrm{S} = S_2\,D^{-7/6}
\end{equation}
for short exposures, and
\begin{equation}
\sigma_\mathrm{L} = S_3\,D^{-2/3}\tau^{-1/2}
\end{equation}
for a long exposure $\tau$.

These expressions follow directly from the Eqs.~(\ref{eq:short_s2}) and (\ref{eq:long_s3}). To convert the values of the Table~\ref{tab:comp2} and \ref{tab:compar} into stellar magnitudes, they should be multiplied by the constant 1.086, i.e., the amplitude of the scintillation noise in magnitudes is $\sigma_\mathrm{mag} = 1.086\,\sigma$.

As we pointed in Sect.~\ref{sec:theory}, for the SE regime the central obscuration effect should be considered. A good approximation is presented in \citet{2012bMNRAS}. For extra-large telescopes ($D \gtrsim 8$~m), the effect of the turbulence outer scale is substantial in both the SE and LE regimes, and should also be taken into account \citep{2012bMNRAS}.

The results shown in these tables are in good agreement with the estimates of scintillation parameters $S_2$ and $S_3$ for Tololo and Pach\'on obtained in \citet{Kenyon2006} by numerical calculation of the moments $\mathcal M_2$ and $\mathcal Y_2$ by Eqs.~(\ref{eq:s2m2}) and (\ref{eq:s3y2}) on the basis of the measured OT vertical profiles and the modelled wind profile. Our method gives for the Maidanak observatory a similar estimate of $S_3 = 0.0030$ \citep{2011AstL}. Unfortunately, other statistically reliable estimates of the scintillation noise are not available to our knowledge.

The results indicate that scintillation noise is mostly defined by the global turbulence at altitudes of 10--15\,km, at the tropopause and above. The only site in our list where the local effects are noticeable and make a difference with other sites is the Tolonchar. In this case, the proximity to the main ridge of the Andes which is quite high and perpendicular to the global air circulation could create a quasi-stationary vertical vortex with a scale of tens of kilometres.

The stronger scintillation noise (about 10\% excess) at the SPM observatory is likely a result of a biased seasonal distribution of the observations together with a significant variability from night to night (see Table~\ref{tab:night3}). On the other hand, the most probable value of $S_3$ at this observatory exceeds the value at Paranal by only 3\%.

\begin{table}
\caption{Comparison of the $S_2$ distribution at all sites (the units are $\mbox{m}^{7/6}$).
\label{tab:comp2}}
\centering
\begin{tabular}{lrrrrrrrr}
\hline\hline
 & \multicolumn{3}{c}{$S_2$ quartiles}\\
 & 25\% & 50\% & 75\% \\
\hline
Armazones & 0.0108 & 0.0132 & 0.0166 \\
La Chira & 0.0127 & 0.0155 & 0.0198 \\
Mauna Kea & 0.0088 & 0.0109 & 0.0139 \\
Pach\'on & 0.0124 & 0.0155 & 0.0197 \\
Paranal & 0.0117 & 0.0144 & 0.0181 \\
S.~Pedro Martir & 0.0117 & 0.0147 & 0.0192 \\
Shatdzhatmaz & 0.0107 & 0.0132 & 0.0168 \\
Tolar & 0.0113 & 0.0136 & 0.0169 \\
Tololo & 0.0122 & 0.0150 & 0.0191 \\
Tolonchar & 0.0116 & 0.0146 & 0.0191 \\
Ventarrones & 0.0120 & 0.0150 & 0.0190 \\
\hline\hline
\end{tabular}
\end{table}

\begin{table}[h]
\caption{Comparison of the $S_3$ distribution at all sites (the units are $\mbox{m}^{2/3}\,\mbox{s}^{1/2}$). \label{tab:compar}}
\centering
\begin{tabular}{lrrrrr}
\hline\hline
 & $S_3$ Mode & \multicolumn{3}{c}{$S_3$ quartiles} \\
 & & 25\% & 50\% & 75\% \\
\hline
Armazones & 0.00300 & 0.00266 & 0.00329 & 0.00410 \\
La Chira & 0.00328 & 0.00294 & 0.00357 & 0.00439 \\
Mauna Kea & 0.00265 & 0.00238 & 0.00290 & 0.00358 \\
Pach\'on & 0.00298 & 0.00265 & 0.00325 & 0.00402 \\
Paranal & 0.00311 & 0.00275 & 0.00336 & 0.00410 \\
S.~Pedro Martir & 0.00310 & 0.00280 & 0.00352 & 0.00453 \\
Shatdzhatmaz & 0.00280 & 0.00251 & 0.00320 & 0.00409 \\
Tolar & 0.00329 & 0.00290 & 0.00346 & 0.00414 \\
Tololo & 0.00303 & 0.00265 & 0.00324 & 0.00396 \\
Tolonchar & 0.00353 & 0.00306 & 0.00386 & 0.00488 \\
Ventarrones & 0.00320 & 0.00272 & 0.00337 & 0.00413 \\
\hline\hline
\end{tabular}
\end{table}

\subsection{Comparison of the S3 parameter with the Young's equation}
\label{sec:compar_s3}

Until now, astronomers \citep[see, e.g.,][]{Everett2001,Mann2011} used the Young's formula \citep{Young1967} to estimates the contribution of scintillation noise to the accuracy of photometric measurements:
\begin{equation}
\sigma_\mathrm{L} = 0.0030\, D^{-2/3} M_\mathrm{z}^{3/2}\,e^{-h_\mathrm{obs}/h_0} \tau^{-1/2},
\label{eq:young}
\end{equation}
where $D$ is telescope diameter in meters, and $h_\mathrm{obs}$ is the observatory altitude above sea level. Since the original expression contains a bandwidth rather than exposure, there was a misunderstanding in the translation of one to another. The (\ref{eq:young}) is taken from \citep{Gilliland1993}, where it was corrected after the intervention of \citet{Young1993}. The dependence on the observatory altitude was originally proposed by \citet{Reiger1963}, who assumed an exponential dependence $C_n^2$ from altitude, but the scale height of $h_0 = 8$~km was established by Young. The numerical coefficient was determined from observations mainly with the 0.9~m telescope.

For comparison with our data, we rewrite the previous formula as $S_3 = 0.0030\,e^{-h_\mathrm{obs}/h_0}$. Using this formula and the site altitudes from Table~\ref{tab:sites}, we obtain $S_3$ estimates from 0.0018 for the highest summit of Mauna Kea to 0.0023 for the Shatdzhatmaz. A comparison with Table~\ref{tab:compar} shows that the Young's formula underestimates the median amplitude of the scintillation noise by a factor of 1.5 (scintillation power by two times). The values of $S_3$ inferred from the Young's formula are similar to the first quartiles of its actual distributions derived both here and in \citet{Kenyon2006,2011AstL}.

The dependence of the scintillation on the observatory altitude is ambiguous because of many factors affecting it. Our results do not show this dependence for the most of studied sites located between 2000 and 3000~m a.s.l. because this effect does not exceed the accuracy of the method. Only for the highest site, Mauna Kea, we can possibly relate lower scintillation noise to higher altitude. On the other hand, the Mt.~Ventarrones data show domination of local effects.

\subsection{Conclusions}

This paper presents the results of the evaluation of scintillation noise in observations on telescopes with large diameter $D \gg r_\mathrm{F}$ or $D \gtrsim 1$~m for the optical and near-infrared in the Earth's atmosphere. These estimates are obtained by an indirect method based on the data of the measurement with the MASS instruments, without involving measurements on large telescopes. However, 1) this method has a reliable theoretical basis, 2) identical instruments were involved and measurements were obtained using the same technique, 3) data for each studied site were obtained over a long time period and have a large statistics.

The scintillation noise at short exposures is quantified by the parameter $S_2$ proposed in \citet{Kenyon2006}, while the parameter $S_3$ characterizes observations with long exposures. Using these parameters, one can calculate the scintillation noise for a telescope of any reasonable diameter for any long exposure. For example, 8-m telescopes such as Gemini or LSST at Cerro Pach\'on will have scintillation noise of 50\,$\mu$mag for $\tau = 5$\,min on a typical night.

The data from seven of the 11 sites studied here should be of particular interest to astronomers because they describe the conditions at the existing observatories Tololo, Pach\'on, Paranal, S.~Pedro Martir, Mauna Kea, or the observatories soon to come into operation (Armazones, Shatdzhatmaz). Scintillation noise measurements at four other summits in the northern Chile are more interesting from the point of view of general turbulence behaviour in the upper atmosphere.

In addition to the general statistical characteristics of the scintillation noise parameters, their temporal variability was investigated, from the short time scales of minutes to seasonal variations. Characteristics of the short-term (during the night) changes are important for optimizing observational strategy in the high-precision photometry. Usually we can assume that the scintillation noise power is stable enough for a $0.5 - 1$~h, but sometimes it may change significantly on timescales of $10 - 15$~min.

The seasonal variations are significant enough and in some observatories they reach a factor of two in power. They are directly related to the latitude of the observatory. The minimum amplitude is observed for Mauna Kea which is close to the equator and the maximum for the mid-latitude observatory at Mt.~Shatdzhatmaz.

The main conclusion from the comparison of scintillation noise at different observatories is that there are no major differences. For the purpose of choosing the best site for high-precision photometry, all sites are essentially equal. A much larger effect can be achieved by choosing the best season for observations and by using the real-time information about the power of the scintillation in operational planning of the photometric observations.

Apart from the high-precision photometry, the parameters measured in the paper can be interesting for error budget evaluation of high-precision differential astrometry \citep{Kenyon2006,Cameron2009}.

\begin{acknowledgements}
The authors thank the Sternberg Astromomical Institute MASS group: N.\,Shatsky, B.\,Safonov, S.\,Potanin, M.\,Kornilov, provided data from Mt.~Shatdzhatmaz. The data for Cerro Tololo and Pachon were obtained from the robotic site monitor operated by the CTIO site testing team. Data from Armazones, Tolar, Tolonchar and Mauna Kea were taken in the framework of the TMT site testing program.
\end{acknowledgements}

\bibliographystyle{aa}

\bibliography{scint_noise}

\appendix

\section{The weighting function approximation}

\label{sec:approxim}

The best approximation is found in the same way as for other atmospheric moments \citep{2003MNRAS}. We solve a linear system of equations ${\rm U^\prime}{\bf d} = {\bf q}$, where ${\rm U^\prime}$ is the weighting function matrix of dimension $m \times n$, ${\bf d}$ is the vector of unknown coefficients, ${\bf q}$ is the desired function $10.66 z^2$. The number of nodes $n = 50$ of the distance grid $\{z_i\}$ is substantially greater than the maximum number of indices $m = 10$. As always, a log-uniform distance grid is used, with higher density at low altitudes.

The system is solved by singular value decomposition, discarding singular values less than $ 5\cdot 10^{-4}$. The quality of the approximation is controlled by the noise amplification factor $F_\mathrm{N} = (\sum_j d_j^2)^{1/2}/\sum_j d_j$. This system is weighted $\propto z^{-1}$ to ensure best fit in the range of $5 - 25$~km. Here we use only the four normal indices because it became clear that an approximation which includes the differential indices is better only at low altitudes, the least interesting zone from the standpoint of the scintillation in large telescopes.

Using the linear relationship between the weighting functions and the scintillation indices, we can finally write
\begin{equation}
S^2_3 = \sum_j d_j s^2_{\mathrm{L}j}, \qquad j = \mathrm{A, B, C, D},
\label{eq:s3sum}
\end{equation}
where $s^2_{\mathrm{L} j}$ are the LE indices (with 1~s exposure) measured in the MASS apertures A, B, C, and D.

The weighting functions $U'(z)$ and, therefore, the coefficients $d_j$, depend on the aperture dimensions and spectral sensitivity of the MASS detectors. Although all MASS/DIMM devices are almost identical, they have been used with different feeding optics and somewhat differ in the spectral response of the detectors and in the magnification coefficient $k$ for overall optical system ``MASS+telescope'' that determines the aperture diameters in plane of the entrance pupil. Therefore, the coefficients $d_j$ were computed for each instrument considered in this work individually. These values are given in Table~\ref{tab:coeff_d}.

\begin{table}
\caption{Coefficients $d_j$ of the approximation of the function  $10.66\,z^2$ by a sum of weighting functions $U'(h)$ for the different MASS/DIMM devices. The devices are marked with their  assembly number and project. The magnification coefficient $k$ and  noise factor $F_\mathrm{N}$ are listed as well. \label{tab:coeff_d}}
\centering
\tabcolsep=4pt
\begin{tabular}{@{}llrrrrrr@{}}
\hline\hline
Site & Device & $k$ &$d_\mathrm{A}$&$d_\mathrm{B}$&$d_\mathrm{C}$&$d_\mathrm{D}$&$F_\mathrm{N}$\\
\hline
Armazones A & MD05 & 15.9 & -0.009 & 0.008 & -0.029 & 0.107 & 1.47 \\
Armazones B & MD31 & 14.0 & -0.014 & 0.020 & -0.054 & 0.117 & 1.89 \\
La Chira & MD21 & 16.1 & -0.009 & 0.006 & -0.024 & 0.104 & 1.40 \\
Mauna Kea & MD08 & 15.8 & -0.010 & 0.008 & -0.031 & 0.108 & 1.48 \\
Pach\'on & MD07 & 14.5 & -0.011 & 0.015 & -0.047 & 0.115 & 1.74 \\
Paranal & LITE & 16.8 & -0.009 & 0.004 & -0.020 & 0.103 & 1.37 \\
S.\,Pedro M\'artir & MD11 & 15.8 & -0.009 & 0.006 & -0.024 & 0.101 & 1.40 \\
Shatdzhatmaz & MD09 & 16.3 & -0.009 & 0.007 & -0.026 & 0.107 & 1.42 \\
Tolar & MD05 & 15.9 & -0.009 & 0.008 & -0.029 & 0.107 & 1.47 \\
Tololo & MD02 & 15.0 & -0.010 & 0.012 & -0.040 & 0.112 & 1.63 \\
Tolonchar & MD02 & 15.0 & -0.010 & 0.012 & -0.040 & 0.112 & 1.63 \\
Ventarrones & MD21 & 16.1 & -0.009 & 0.006 & -0.024 & 0.104 & 1.40 \\
\hline\hline
\end{tabular}
\end{table}

\section{Filtering of the LE scintillation indices}
\label{sec:filtering}

During the computation of the indices using the Eq.~(\ref{eq:sdif}) and the data stored in the file {\tt *.stm}, a set of the flux values $\{F_i \}$ is omitted if 1) there is no information about the sky background or the background is too high, 2) the flux ratio between MASS apertures B and C does not satisfy the condition $F_\mathrm{B} < F_\mathrm{C}$, 3) the number of 1-s flux values in a 1-min. accumulation period $N<10$, and 4) the flux in aperture D is low, $F_\mathrm{D} < 100$. These criteria eliminate obviously wrong or inaccurate data. However, they do not exclude all situations with incorrect data and we developed additional data-filtering criteria:
\begin{itemize}
\item $g_1$ is the ratio of the maximum term in eq.~(\ref{eq:sdif}) to  the total sum for aperture D. This parameter detects strong outliers.
\item $g_2$ the ratio of the flux range in the aperture D between its  95\% and 5\% quantiles (i.e. ignoring the 2 lowest and 2 largest  values in series of 60 points) to its rms fluctuations $\sqrt{s^2}$. This indicator is sensitive to a trend.
\item $g_3$ is calculated similarly $g_2$ using the full flux range. It monitors isolated overshoots or, in their absence, the  general trend.
\end{itemize}
We determined the filtering thresholds for these indicators from their empirical distributions.

\begin{figure}
\centering
\psfig{figure=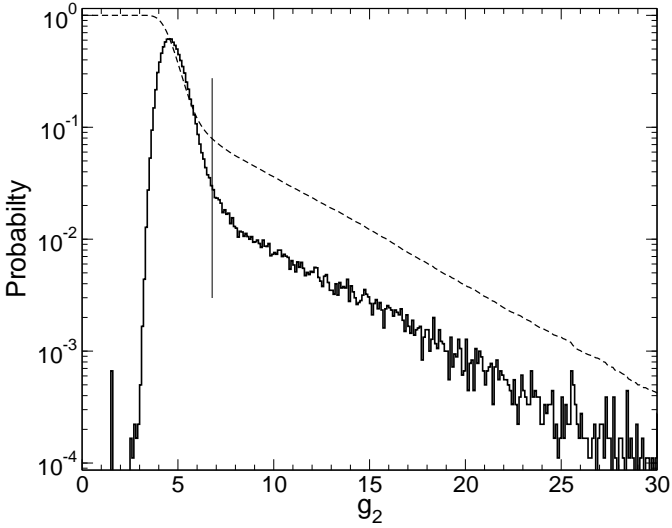,height=9cm,angle=-90}
\caption{
Distribution of the $g_2$ indicator for all data from Shatdzhatmaz site. The dashed line represents the complementary cumulative distribution. The thin vertical line shows the chosen cut-off threshold. \label{fig:g2distr}}
\end{figure}

As an example, the differential and complementary cumulative distributions of $g_2$ for the data obtained at the Shatdzhatmaz are shown in Fig.~\ref{fig:g2distr}. Similar distributions are found at other sites. The inflection point on the right (descending) branch of the differential curve is chosen as the threshold for all sites (at larger $g_2$ the distributions clearly change their character).  In this example, the cut-off is set at $g_2 > 6.7$, which discards $\approx 8\%$ of the data. Visual inspection confirms that the discarded data are affected by either clouds or large tracking errors.

Using differences for calculating the LE scintillation indices does not completely suppresses the low-frequency flux variation, increasing the estimated variance. Most of these variations have the form of a trend during the accumulation time. We denote the relative change in the flux due to the trend by $\Delta $. It is easy to show that a trend adds $\Delta^2/2N^2$ to the true value of the LE index $s^2_\mathrm{L}$. If the relative error of the index estimate must not exceed $p$, the following condition must be satisfied:
\begin{equation}
\Delta < N\,(2 p\, s^2_\mathrm{L})^{1/2}.
\label{eq:trend}
\end{equation}

If $p = 0.01$ and $N = 60$, the allowable trend must be $\Delta < 8.5\,(s^2_\mathrm{L})^{1/2}$. Since the LE indices in the apertures A, B and C enter in the result (the parameter $S^2_3$) with smaller coefficients, and are themselves larger than the index in D, it is sufficient to check the condition (\ref{eq:trend}) for the aperture D only. In addition, monitoring of the absolute value of the trend is also needed, since large systematic changes in flux induce such an increase in the variance that the condition (\ref{eq:trend}) again begins to be fulfilled.

We used yet another powerful criterion based on the theoretical relation between the measured values. The behaviour of the weighting functions $U'(z)$ restricts the ratios of the LE indices. Clearly, that the ratios $U'(z)$ for the apertures A, B, and C, to the weighting function for the aperture D not equal to the ratio of the indices themselves, but for any profiles of OT and wind the ratios of $s^2_\mathrm{L}$ must lay between the theoretical minimum and maximum.

\begin{figure}
\centering
\psfig{figure=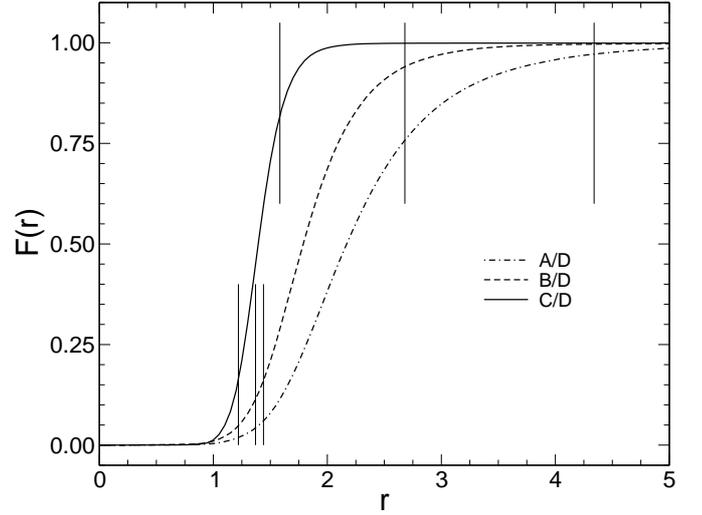,height=9cm,angle=-90}
\caption{ Distributions of the ratios of the $s^2_\mathrm{L}$ indices in the apertures A, B, C to the index in D measured at Mauna Kea. Vertical thin lines indicate the theoretical ratio for the distance $z= 2$~km (top) and 32~km (bottom) for the C/D, B/D and A/D, from left to right.
\label{fig:eratio}}
\end{figure}

The ratio of LE indices $s^2_\mathrm{L,C}/ s^2_\mathrm{L,D} $ is weakly dependent on the distance, which is reflected in the form of the observed distribution of the ratio of $s^2_\mathrm{L}$ in the apertures C and D shown in Fig.~\ref{fig:eratio}. The cumulative distribution is very steep and the differential distribution is very narrow. The ratio of indices in the apertures A and D typically has a broader distribution, which is explained by the behaviour of the functions $U'(z)$. This ratio is most sensitive to the specific characteristics of the MASS device and to the minimum altitude of the turbulence which produces noticeable scintillation. The minimum ratio of LE indices is reached when all the turbulence is at $\sim 32$~km.

Fig.~\ref{fig:eratio} indicates a good match between the measured ratios and their theoretical estimates, so we boldly used the experimental distributions to determine the lower and upper limits (approximately at the level of 0.1\%) for rejecting the outliers. Typical instrumental causes of such outliers are a wrong sky background value or an accidental partial vignetting of the entrance pupil. Of course, random errors in the LE indices also widen the observed distribution.

\section{Error estimation of LE indices}

The errors of LE indices $s^2_\mathrm{L}$ were estimated in two ways. First of all, assuming stationarity over one minute, normal distribution, and uncorrelated 1-s fluxes, we estimated the probable relative error of the sample variance $\varepsilon = (2/N)^{1/2}$ as 0.18 for all apertures.

The second method consists in computing the mean squared difference between the logarithms of successive estimates: $\delta_i = (\ln s^2_i - \ln s^2_{i+1})^2$. Assuming a quasi-stationarity, the relative error is $\varepsilon = (0.5 \langle \delta \rangle)^{1/2}$. The probability distribution of $\delta$ is similar to the $\chi$ distribution with one degree of freedom. The medians of such estimated errors are thus 0.175, 0.174, 0.172 and 0.172 for the apertures A, B, C and D (for the Shatdzhatmaz data).

Assuming the same relative errors of $s^2_\mathrm{L}$ for all apertures, $\varepsilon = 0.18$, and assuming uncorrelated errors in all apertures, we calculated the relative error $E_1$ of the parameter $S^2_3$. Because noise amplification by the linear combination (\ref{eq:s3sum}) depends on the coefficients which differ between individual MASS instruments, the errors of $S^2_3$ are also different. The median values of these errors are $0.28 - 0.36$.

The errors in the different apertures are partially correlated \citep[see][]{2011ExA}, so the estimate $E_1$ of the errors was additionally checked by calculating the mean difference of the logarithms of the adjacent $S^2_3$ values, as for the LE indices. These estimates $E_2$ are somewhat smaller than $E_1$, but there is a good agreement between them.

The average errors of the 4-minute $S^2_3$ estimates obtained in the this way are listed in Table~\ref{tab:s3}. It can be seen that they are practically identical for all sites. The same method was used to estimate the errors of the parameter $S^2_2$, which are typically $0.06 - 0.09$ (see Table~\ref{tab:s2}). These errors are smaller because the estimates are calculated from $\sim 1000$ samples.

\end{document}